\documentclass{osa-article}

\journal{osac}

\articletype{Research Article}

\usepackage{lineno}
\usepackage{caption}
\usepackage{subcaption}

\usepackage{bbold}
\begin{document}

\title{Graph model for multiple scattering in lithium niobate on insulator integrated photonic networks}

\author{Xiyue Sissi Wang,\authormark{1*} Romolo Savo,\authormark{1,2}, Andreas Maeder,\authormark{1} Fabian Kaufmann,\authormark{1} Jost Kellner,\authormark{1} Andrea Morandi,\authormark{1} Stefan Rotter,\authormark{3} Riccardo Sapienza,\authormark{4} and Rachel Grange\authormark{1}}

\address{\authormark{1} ETH Zurich, Department of
Physics, Institute for Quantum Electronics, Optical
Nanomaterial Group, 8093 Zurich, Switzerland\\
\authormark{2}Centro Ricerche Enrico Fermi (CREF), Via Panisperna 89a, Rome 00184, Italy\\
\authormark{3}Institute for Theoretical Physics, Vienna University of Technology (TU Wien), Vienna, Austria\\
\authormark{4}The Blackett Laboratory, Department of Physics,
Imperial College London, London SW7 2BW, United Kingdom\\
}

\email{\authormark{*}siswang@phys.ethz.ch} 



\begin{abstract}
We present a graph-based model for multiple scattering of light in integrated lithium niobate on insulator (LNOI) networks, which describes an open network of single-mode integrated waveguides with tunable scattering at the network nodes. 
We first validate the model at small scale with experimental LNOI resonator devices and show consistent agreement between simulated and measured spectral data. Then, the model is used to demonstrate a novel platform for on-chip multiple scattering in large-scale optical networks up to few hundred nodes, with tunable scattering behaviour and tailored disorder. Combining our simple graph-based model with material properties of LNOI, this platform creates new opportunities to control randomness in large optical networks.
\end{abstract}

\section{Introduction}
Graphs, or networks, are commonly used structures for modelling physical systems. The utility of their geometry is twofold: on the one hand, complex dynamics can be modelled by a large network of simple elements and their pairwise connections. Such ideas have been applied to numerous examples ranging from quantum information transmittance\cite{quantumRandomNet}, improving power grid robustness \cite{powerGrid}, and community structures in biological and social systems \cite{graphCommunity}.  In the context of optics, networks describe optical circuits, which can perform elaborate operations using connections between simple optical elements \cite{opticalCircuitReview,opticalFPGA,QOcircuits1,QOcuircuits2}.
On the other hand, a graph often serves as an appropriate model of dimension reduction and discretisation of a real physical medium \cite{Kuchment2002_thinStructures}, for example for computing the band structure of graphene \cite{graphene} or carbon nanotubes \cite{carbonNanotube}.

The second case above often falls within the theory of quantum graphs \cite{QGbook}. Quantum graph models may not necessarily describe systems which are quantum mechanical in nature; for our purposes, they describe waves (where interference comes into play) on metric graphs (where both the topology of connections and geometry of distances play a role). Classical realisations for quantum graphs have been achieved in the microwave domains, using networks of coaxial cables \cite{microwave2004,microwave2014}. For optics, such models have been applied to random lasers: in these examples, the graph edges are either composed of dye doped polymer nanofibres spun on chip \cite{randLaser_Gaio2019,randLaser_Saxena2022} or optical fibres with coherent optical amplifying sections as gain \cite{LANER}.

In this work, we apply network-based models to linear photonic integrated circuits (PICs), as a platform for studying on-chip multiple scattering with fully controlled and pre-designed disorder. To this end, several modifications are made compared to existing quantum graph models for optical systems. For one, focus is put on the ability to fully specify the scattering behaviour at each scattering site in the network, which allows for greater flexibility and better agreement with integrated optical networks compared to Neumann-type continuity conditions commonly used in the quantum graph literature \cite{QGbook,QGreview1} and in previous work on integrated random laser \cite{randLaser_Gaio2019,randLaser_Saxena2022}. Secondly, our model focuses on open networks, i,e. the linear transport properties by looking at the continuous spectra, rather than discrete resonance modes. While closely related, the coupling to open ports modifies the resonances of a network, and the open model matches more closely with the physical measurements. It is also worth noting that this approach differs from traditional optical circuits \cite{opticalCircuitReview,opticalFPGA,QOcircuits1,QOcuircuits2}, since backscattering is utilised as a source for generating interference.

Small scale graphs, which consist of a few to several dozens nodes, can be constructed to model traditional integrated optical devices such as ring resonators, and automatise the calculation for circuits composed of several coupled elements. We demonstrate that by taking into account the appropriate material properties, quantitative agreement is reached between simulated and measured spectra for integrated lithium niobate on insulator (LNOI) devices. LNOI is chosen as the physical platform due to its electro-optic (EO) property, allowing tuning of network geometry by applying voltage across waveguides \cite{LNOIreview_Loncar,Marc_EOtuning}. EO tuning comes into play when we look at scaled up network models, where complex random networks serve as a model for multiple light scattering \cite{Mosk2012}. In such cases, tuning allows for control over scattering and propagation parameters on individual optical elements.

By implementing a scattering matrix model which is significantly simplified compared to numerical techniques such as finite-difference time domain (FDTD) and finite element method (FEM), we are able to explore integrated photonic networks with few hundred nodes as a new platform for studying multiple scattering. The nature of top-down design and fabrication allows one to control the exact type of randomness introduced into the network, as well as behaviour of individual scatterers. 
 






    

\section{The scattering matrix model}\label{sec:model}
We start with the description of the scattering matrix model for integrated networks, distinguishing between two cases: 1) the closed network, which is the common model adopted for fibre lasing networks \cite{LANER} and quantum graphs \cite{QGbook,QGreview1}, and 2) the open network, which more realistically accounts for the coupling geometry in a measurement setup.

\subsection{Closed network model and the secular equation}
\begin{figure}[!h]
     \begin{minipage}{0.28\textwidth}
     
    \vspace{20pt}
        \begin{subfigure}[b]{\textwidth}
                 \includegraphics[width=\textwidth]{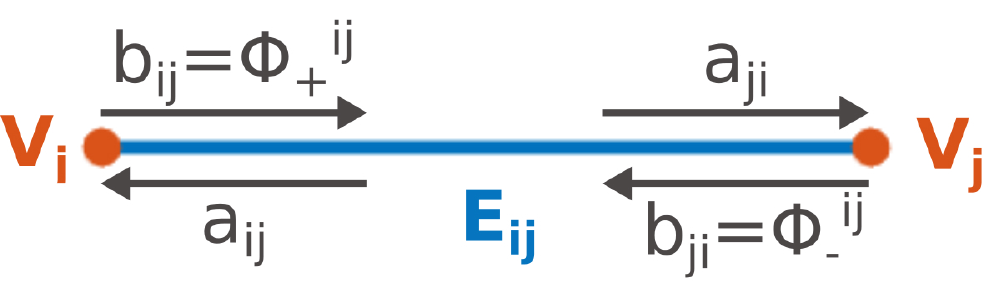}
                 \caption{}
                 \label{fig:demo_edge}
        \end{subfigure}
        \begin{subfigure}[b]{\textwidth}
                 \includegraphics[width=\textwidth]{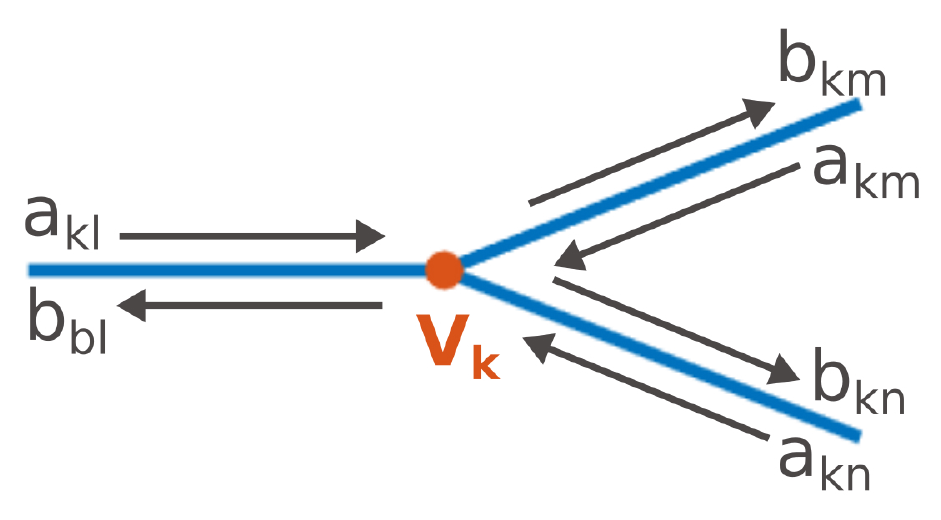}
                 \vspace*{-15pt}
                 \subcaption{}
                 \vspace*{15pt}
                 \label{fig:demo_node}
        \end{subfigure}
    \end{minipage}
    \vspace*{5pt}
     \begin{minipage}{0.33\textwidth}
\begin{subfigure}[b]{\textwidth}
         \includegraphics[trim={1cm 0 1cm 0},clip,width=\textwidth]{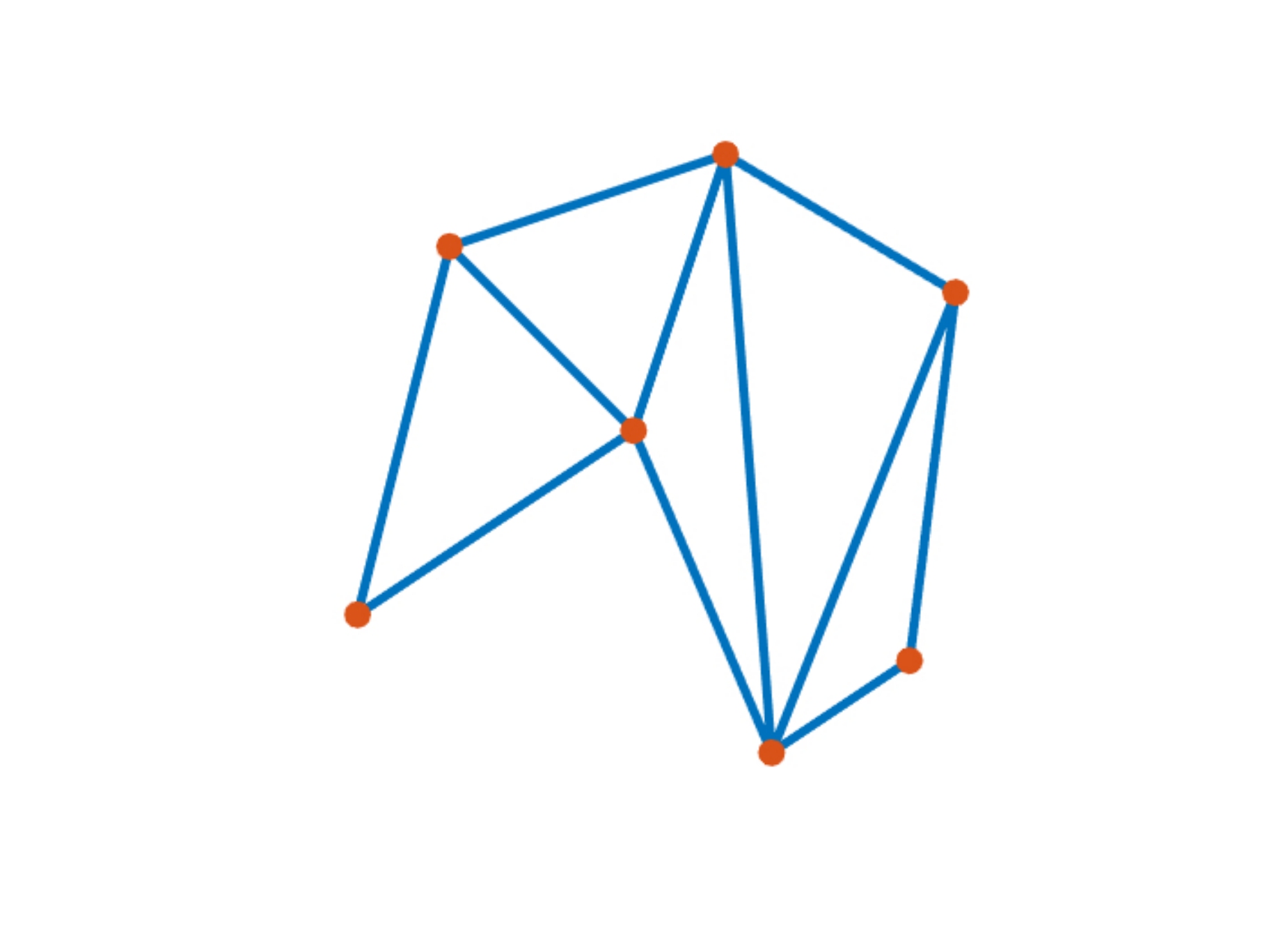}
         \vspace{-30pt}
         \hfill \hspace*{1cm}
         \caption{}
          \vspace{30pt}


         \label{fig:demo_closed}
\end{subfigure}
\end{minipage}
\begin{minipage}{0.33\textwidth}
\begin{subfigure}[b]{\textwidth}
         \includegraphics[trim={1cm 0 1cm 0},clip,width=1.0\textwidth]{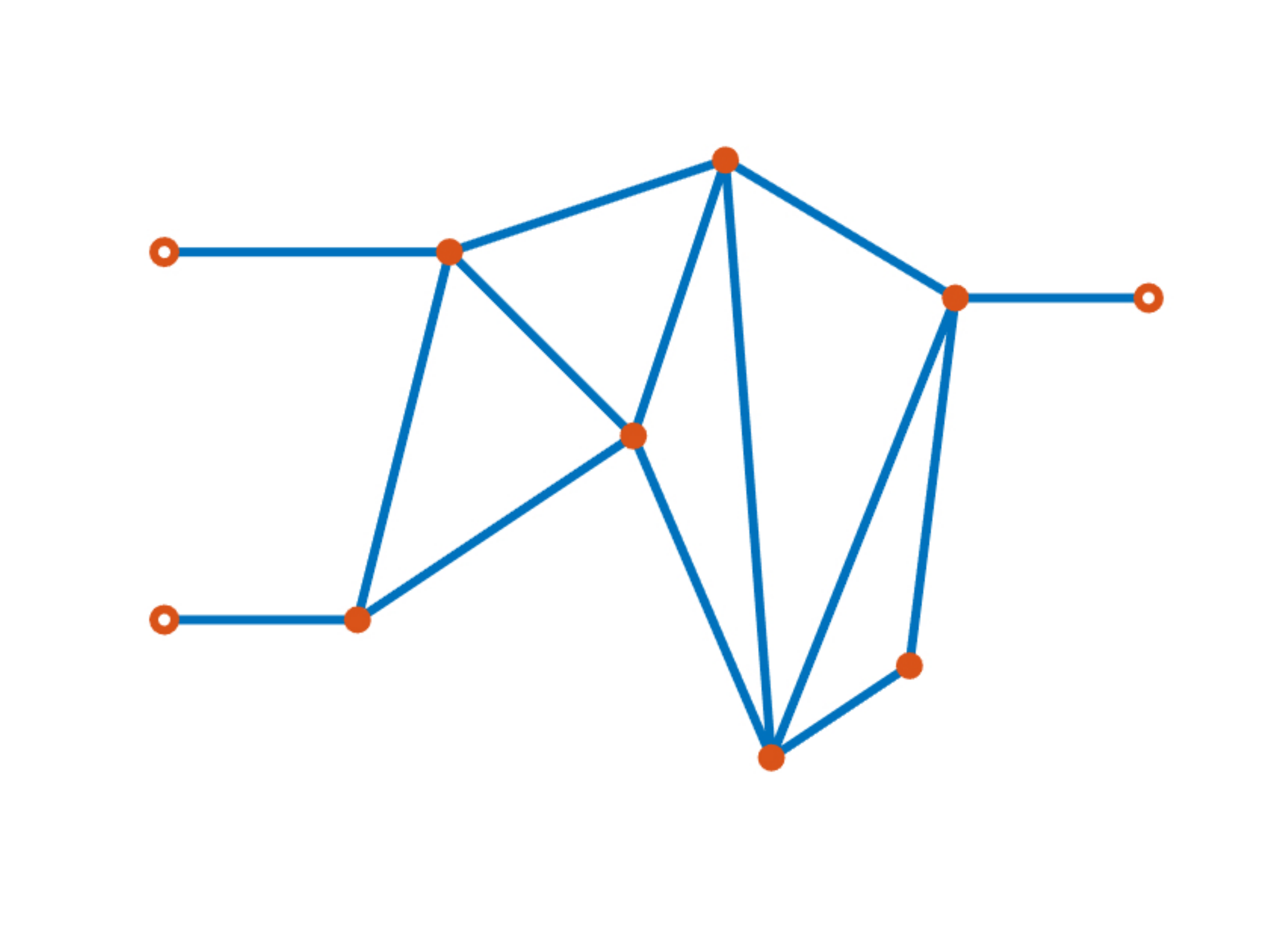}
         \vspace{-25pt}
         \hfill \hspace*{1cm}
         \caption{}
          \vspace{25pt}
         \label{fig:demo_open}
\end{subfigure}
\end{minipage}
\vspace{-25pt}
        \caption{Network components in the scattering matrix model. (a) Amplitudes along an edge are related by the propagation matrix $A=PB$. Note that although they are equal in magnitude, a coordinate change exists between the propatation for $b_{ji}$ and $\Phi_{-}^{ij}$, as evidenced by their index orders. (b) Amplitudes at a node are related by the scattering matrix $B=SA$. (c) An example for a closed network, with nodes of degrees 2, 3, and 4. (d) The same network becomes an open network after attaching leads/external nodes. External nodes, shown with hollow markers, are where one sets boundary conditions corresponding to experimental input and measures the output.}
        \label{fig:demo_networks}
\end{figure}

The integrated network is represented by a mathematical graph $G=(E,V)$, defined by a set of edges $E$ and a set of vertices $V$ (Fig.~\ref{fig:demo_networks}). For our purposes, the graph is assumed to be connected (i.e. there exists no isolated components) and undirected (i.e. the edges have no inherent orientation). The edges, or links, correspond to single-mode waveguides along which light propagates. The vertices, or nodes, correspond to the scattering elements in the network: they can be sites of reflection and power splitting, for example Bragg gratings and beam splitters. Note that we are restricting our system to 2D, where only in-plane scattering is considered; out-of-plane scattering is incorporated as loss. For this paper, we will use \textit{nodes} to emphasise the fact that scattering sites are scattering elements with internal structure, and \textit{edges} to highlight the metric properties beyond connectivity. Light propagation in the network is modelled as a superposition of two counter-propagating waves with amplitudes $\Phi_+$ and $\Phi_-$, which are defined locally on each edge as
\begin{equation} \label{eq:1}
\Phi(x) = \Phi_{+}e^{ikx} + \Phi_{-}e^{ik(L-x)};
\end{equation}
where $x\in [0,L]$ is the local spatial coordinate defined independently on each edge, $L$ being the length of the edge. The set of amplitudes on all edges fully characterise the field distribution in a graph, and are computed using matrix equations as will be shown below. Edges are joined together by nodes, which introduce further phase difference and local interference effects that are encapsulated in the local scattering matrix for each node.


A solution, or mode, on a graph is an admissible set of amplitudes, which are consistent with the scattering behaviour and geometry of the network. It is obtained by treating scattering (nodes) and propagation (edges) in conjunction through the matrix equations described below. 

To construct the relevant matrices, we define two sets of field amplitudes: $A$, the set of all incoming amplitudes at all the nodes, and $B$, the set of all outgoing amplitudes. Each set is related to $\Phi_+$ and $\Phi_-$ by a simple phase factor, illustrated in Fig.~\ref{fig:demo_edge}. The local scattering matrix at each node is defined by the optical element that it corresponds to; it is a $d\times d$ matrix that maps $A$ to $B$, where the degree $d$ is the number of edges attached to the node. The propagation matrix maps $B$ to $A$ along each edge, and has the form $\begin{pmatrix}a_{ij} \\ a_{ji} \end{pmatrix} = \begin{pmatrix}0 & e^{ikL} \\ e^{ikL} & 0 \end{pmatrix} \begin{pmatrix}b_{ij} \\ b_{ji} \end{pmatrix}$ for an edge of length $L$ between nodes $i$ and $j$. The overall scattering matrix $S$ and propagation matrix $P$ for an entire network is constructed by embedding local matrix elements at the correct index. It is clear in the edge-centred picture that either $A$ or $B$ alone is sufficient for describing a solution (i.e. a full set of amplitudes) in the network, with a set of phase differences between them. One small but important difference between these solutions and $\Phi$ (as specified in Eq. \ref{eq:1}) is the direction: propagating along an edge from $B$ to $A$ always picks up a phase in the same sign $e^{ikL}$, while $\Phi_+$ and $\Phi_-$ pick up $e^{ikL}$ and $e^{-ikL}$, respectively, corresponding to a coordinate change $x\rightarrow L-x$ implicitly performed for the $\Phi_-$ amplitudes.


A solution on the graph is then an eigenvector $A$ with eigenvalue 1 when mapped back onto itself via $PS$:
\begin{equation} \label{eq:eigenvec}
(PS-\mathbb{1})A = 0
\end{equation}
(or alternatively, $(SP-\mathbb{1})B=0$). In analogy to the propagator in the quantum graph context \cite{quantumGraphology}, we define $U(k):=P(k)S$, where we make explicit that the k-dependence comes from the propagation along the edges, while the scattering at the node can be treated as wavelength independent. $U(k)$ is unitary when both $P$ and $S$ are lossless. The modes of the network then correspond to the nontrivial solutions of the secular equation
\begin{equation} \label{eq:secular}
\det(U(k)-\mathbb{1}) = 0 ,
\end{equation}
and may only exist for a discrete set of ${k_s}$.

\subsection{Open network model and the transmission matrix} \label{section:openGraph}
The distinction between \textit{open} and \textit{closed} networks can be expressed as follows: networks with open edges, i.e. edges connected to nodes of degree $1$, are classified as open; otherwise they are closed (Fig.~\ref{fig:demo_closed}-d). The reason for this distinction is that the above characteristic equation formalism poses boundary conditions that are too restrictive for a physical device. 
Consider an external node with degree 1: the scattering matrix for this node is simply complex scalar, characterising the reflection back into the network from the boundary. In particular, if we want external edges to correspond to physical waveguides with no back-reflection, this scalar would be zero, resulting in a solution with zero field on the edge. The constraint matrix would then have the block matrix form 
\begin{equation} \label{eq:open(U-1)}
    PS-\mathbb{1} = \left(
    \begin{array}{c|c}
      -\mathbb{1}_m & 0\\
      \hline
      M_L & M_R
    \end{array}
    \right)
\end{equation}
where $\mathbb{1}_m$ is the $m\times m$ identity matrix, $m$ being the number of external edges; $M_L$ and $M_R$ denote the left and right sections of the remaining $2N_e-m$ rows, $N_e$ being the total number of edges. In this basis, the solution vector is arranged to have the form $A=(A_{in},A_{out},A_{int})^T$, where $A_{in}$ are the amplitudes on external edges travelling into the network, $A_{out}$ are those on external edges coming out of the network, and $A_{int}$ are amplitudes on internal edges.

It is clear that no matrix of this form would satisfy Eq.~\ref{eq:eigenvec} with nonzero $A_{in}$. Instead, we modify Eq.~\ref{eq:eigenvec} to allow for nonzero input: $A = (PS)A + (A_{in},0,0)^T$, and therefore $A = -(PS-\mathbb{1})^{-1}(A_{in},0,0)^T$. Using standard block matrix inversion formulas, we can pick out the relevant submatrix in $-(PS-\mathbb{1})^{-1}$:
\begin{equation} \label{eq:Stotal}
\begin{pmatrix}
    A_{out} \\
    A_{int} 
    \end{pmatrix} = (-M_R^{-1}M_L) A_{in}.
\end{equation}
The matrix $(-M_R^{-1}M_L)$ takes the input amplitudes into the entire network, and maps to all other amplitudes constituting a solution. In particular, we define the matrix $S_{io}$ as the first $m$ rows of $(-M_R^{-1}M_L)$, characterising the input-output relation in the external waveguides of the network via $A_{out} = S_{io} A_{in}$. Under this construction one obtains a continuous spectrum in $k$, the resonances of which correspond to the real part of the discrete eigenvalues of $U(k)$ for the closed network \cite{QGopen}.

One final caveat is that $A_{in}$ differs from the true experimental input $B_{in}$ by the phase picked up from travelling along external edges. In Fig.~\ref{fig:demo_edge}, if $V_i$ is the external node, this is the difference between $a_{ji}$ and $b_{ij}$. We therefore redefine $S_{io} \rightarrow S_{io}P_{ext}$, where $P_{ext}$ is the propagation submatrix applied to the external edges (if one starts by solving for $B$, a similar correction is needed to go from $B_{out}$ to $A_{out}$). This ensures the final $S_{io}$ matrix is reciprocal in both amplitude and phase.

\section{Experimental verification on integrated lithium niobate resonators} \label{sec:smallGraphs}
To verify the model's capability to describe real integrated photonic networks,  simulation results are compared with measurement data on LNOI components fabricated on chip. We demonstrate agreement between the simulations and measurements when material characteristics are accounted for in the model. In particular, we look at two classes of resonators: ring resonators and interferometric resonators. They correspond to two simple types of networks in our formalism. Indeed, a complex network can be interpreted as a complex set of coupled resonators, and these simple resonators can be viewed as the building blocks of larger systems.

The devices were fabricated on a 600~nm X-cut LNOI chip patterned using electron-beam lithography and argon ion milling. To expose the end-facets of the waveguides, the chip was diced. Polarisation maintaining lensed fibers were used to couple to the fundamental transverse electric (TE) mode of the waveguide. The spectral measurement on ring resonators use a tunable laser source and a high dynamic range power meter, while that on interferometric resonators use a broadband superluminescent diode source and an optical spectrum analyser. Details on the fabrication and characterisation techniques are presented in previous work \cite{David_bragg}.

\subsection{Ring resonators}\label{section:RR}
\begin{figure}[!h]
     \begin{minipage}{0.3\textwidth}
\begin{subfigure}[b]{\textwidth}
         \includegraphics[width=0.8\textwidth]{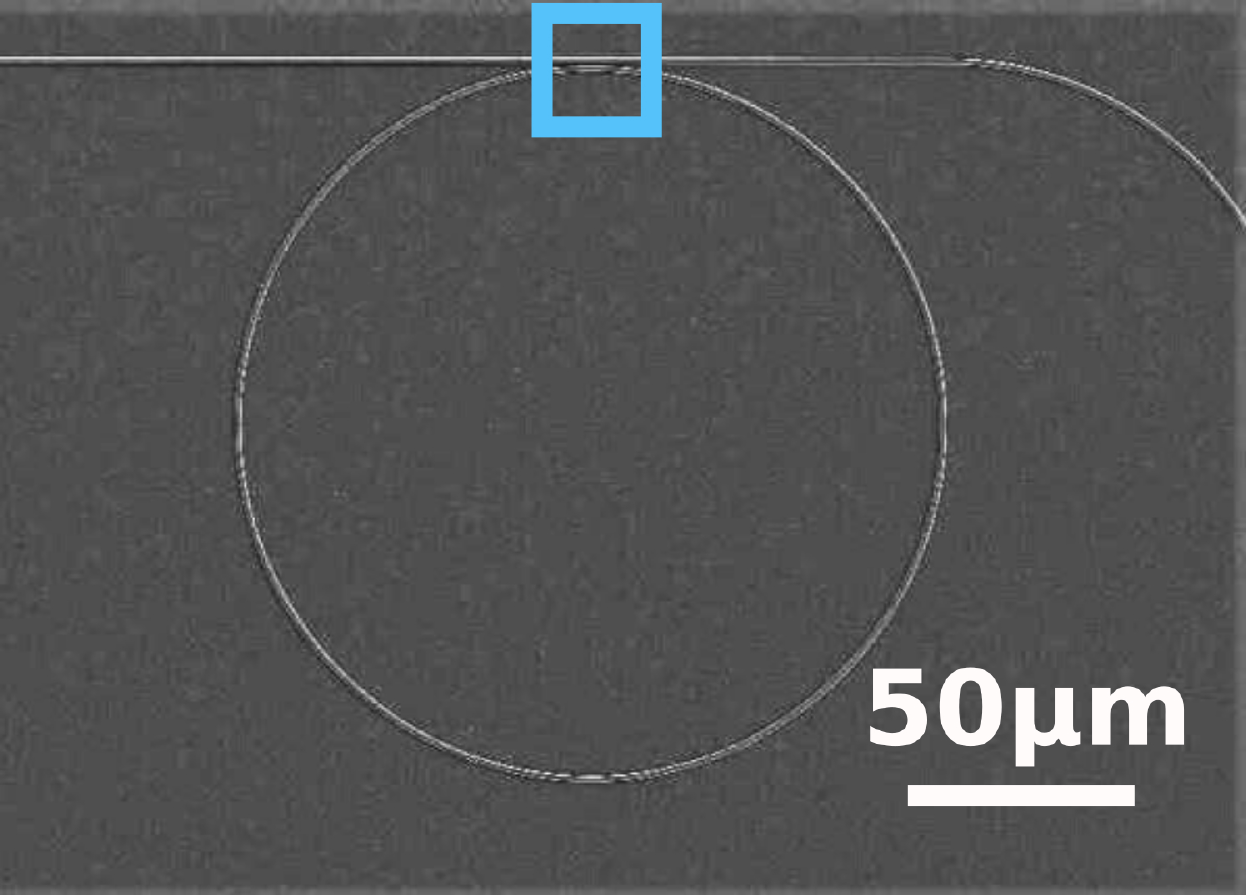}
         \caption{}
         \label{fig:RR_SEM}
\end{subfigure}
\begin{subfigure}[b]{\textwidth}
         \includegraphics[width=0.8\textwidth]{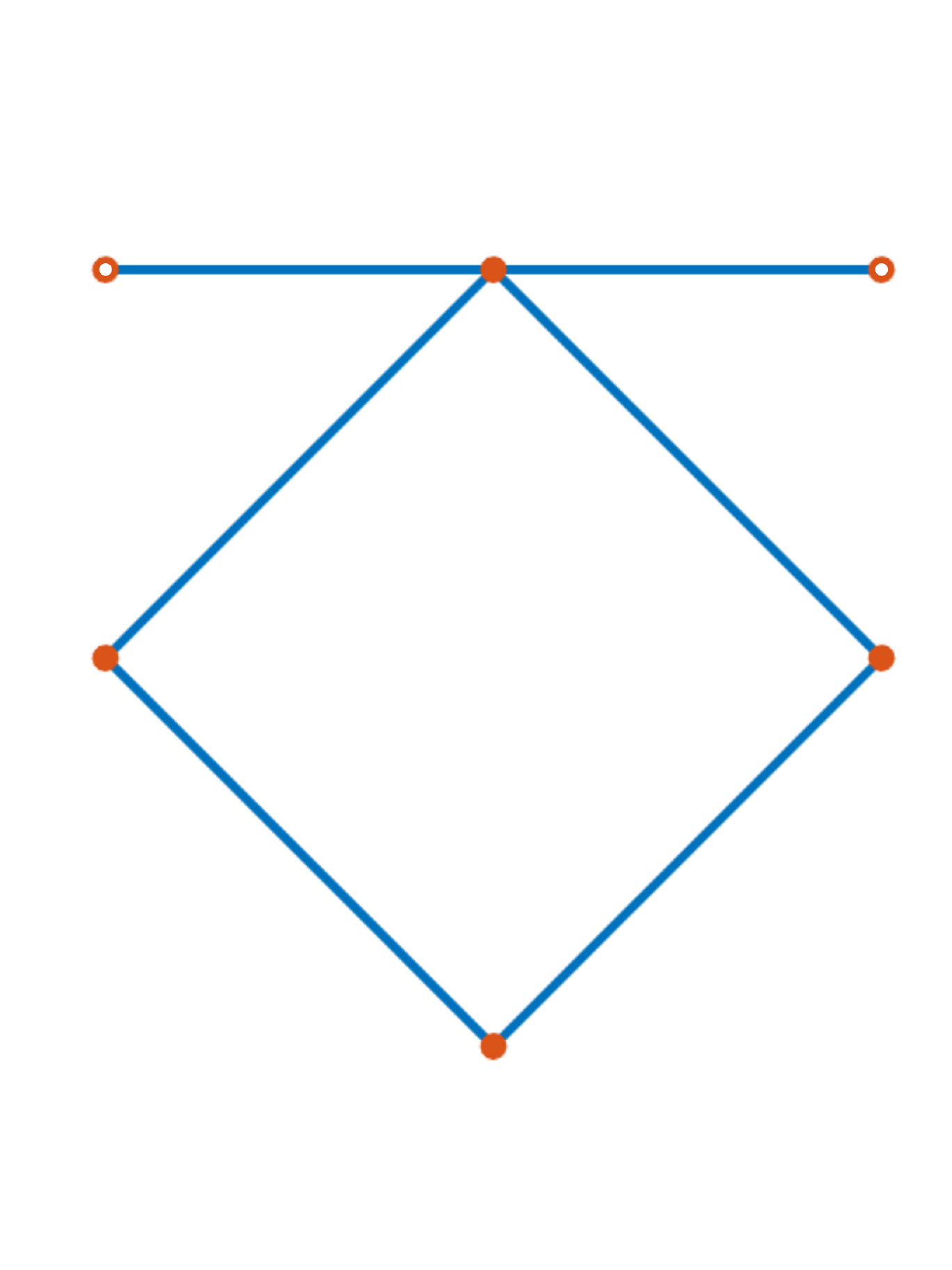}
         \vspace*{-20pt}
         \caption{}
         \vspace*{20pt}
         \label{fig:RR_scheme}
\end{subfigure}
    \end{minipage}%
     \begin{minipage}{0.7\textwidth}
\begin{subfigure}[b]{\textwidth}
         \includegraphics[width=1.0\textwidth]{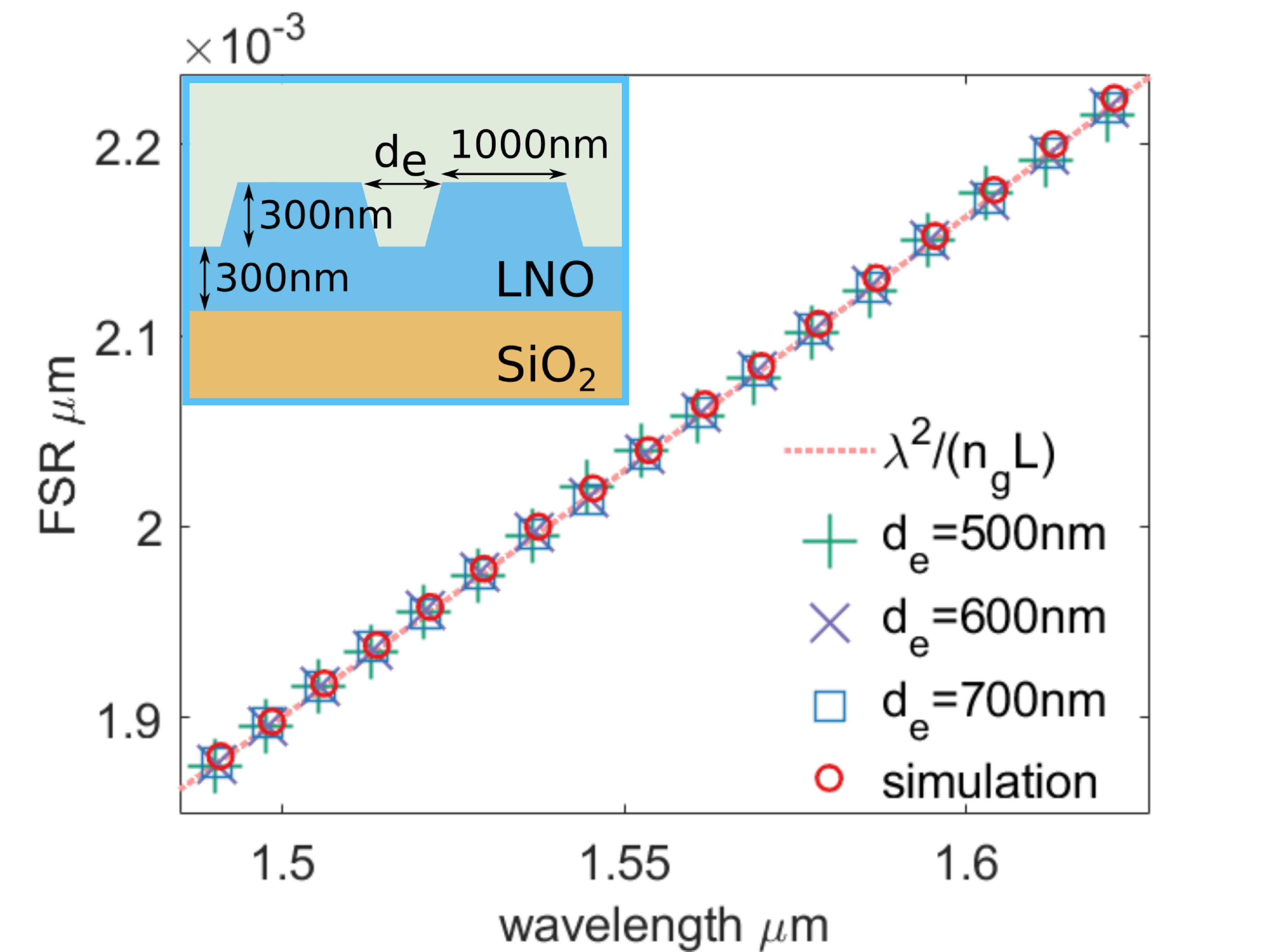}
         \vspace*{-20pt}
         \caption{}
         \vspace*{20pt}
         \label{fig:FSRcompare}
\end{subfigure}
\end{minipage}
        \vspace*{-20pt}
        \caption{Experimental verification on a ring resonator. (a) SEM image of a notch type ring resonator structure fabricated on LNOI thin film. The box indicates the coupling region. (b) Graph used to simulate the ring resonator with the model described in Section \ref{sec:model}. The ring could be represented with one curved edge, but is here shown as square to demonstrate the model does not have to match exact geometry of physical networks, as long as appropriate dimensions and scattering matrices are used.  (c) Free spectral range (FSR) for the single ring resonator, theory (red dotted line) vs. measurement vs. simulation. Inset shows cross section at the coupling region labeled in (a): 300nm etched on 600nm thin film, top width 1000nm, $d_e$ the distance between the bus waveguide and the ring structure. The resonances are unaffected by the evanescent coupling strength as expected, and robustness across devices indicates consistent fabrication quality. For the sake of clarity in presentation, the FSR for one in every four resonances is depicted.}
        \label{fig:1ring}
\end{figure}

A single ring resonator, coupled to one bus waveguide via evanescent coupling, is a notch-type filter commonly found in integrated photonic devices \cite{RRbook_Rabus2007}. Figure \ref{fig:RR_SEM} shows one such device, with \ref{fig:RR_scheme} being the corresponding network. Note the ring can be modelled with one curved edge, but is here shown as a square, demonstrating that a desired balance between model size (number of edges) and abstraction (all straight edges, ease of plotting) can be achieved by properly choosing the scattering matrices. The top center node, corresponding to the evanescent coupler, is characterised by $s_4=\begin{pmatrix}
    0 & 0 & r & it \\ 0 & 0 & it & r \\ r & it & 0 & 0 \\it & r & 0 & 0 \end{pmatrix}$, while the other connecting nodes on the ring simply have $s_2=\begin{pmatrix}
    0 & 1\\ 1 & 0
    \end{pmatrix}$. We assume $|r|^2+|t|^2=1$ for a lossless scattering node.


One can then follow the procedure described in Section \ref{section:openGraph}, and recover the symmetric transmission between the two open ports (up to an additional phase picked up by traveling along the external bus waveguide) to be $T_{16}=(1-re^{-ikC})/(r-e^{-ikC})$, $C$ being the circumference of the ring resonator, in agreement with the analytical solution for notch-type filters \cite{RRbook_Rabus2007}.

In order to reach agreement between simulation results and physical measurements, we define $k_{\textrm{eff}} = k/n_{\textrm{eff}}(\lambda,\theta)$, scaling the vacuum wave number $k$ by the effective refractive index $n_{\textrm{eff}}(\lambda,\theta)$. $n_{\textrm{eff}}(\lambda,\theta)$ accounts for the waveguide geometry (as opposed to bulk lithium niobate), as well as its dispersion ($\lambda$-dependence), and is calculated using Lumerical MODE software. Birefringence of LNOI material is included by the parameter $\theta$, the angle between a waveguide and the ordinary crystal axis on an x-cut lithium niobate thin film. The $\theta$-dependence of $n_{\textrm{eff}}$ is calculated using finite element method, from which the index ellipsoid is obtained. 

For the ring resonator, we assume $n_{\textrm{eff}}$ is uniform across the ring, and the value that it takes is the angle-average $n_{\textrm{avg}}(\lambda) = \frac{2}{\pi}\int_0^{\pi/2}n_{\textrm{eff}}(\lambda,\theta)d\theta$. Furthermore, a small imaginary part $Im(k_{\textrm{eff}}) = 2.3\times10^{-6} \mu m^{-1}$ is included to account for the estimated propagation loss of $0.1~dB/c m$. 

The free spectral range (FSR) is used to quantify the transmission spectra of ring resonators for comparison.  In Fig.~\ref{fig:FSRcompare}, the FSR simulated using the scattering matrix model is compared against measurement data from several physical devices, as well as the theoretical value (FSR$= \lambda^2/n_gL$, $n_g$ the group index). Within the 1.49-1.63$~\mu$m wavelength window, the angle-averaged $n_{\textrm{eff}}$ is fitted by the polynomial $n_{avg}(\lambda) = 0.0006\lambda^2 -0.2652 \lambda + 2.3363$, with wavelengths in $\mu m$ units. 

Excellent agreement is reached between simulation results and several devices with different coupling strengths between the ring resonator and the bus waveguide (Fig.~\ref{fig:FSRcompare}), which verifies the resonance behaviour to be independent of coupling strength, and demonstrates good fabrication quality consistent across devices. Agreement with the theoretical value (where $n_g$=$n - \lambda\partial{n}/\partial{\lambda}$ is computed with the $n = n_{\textrm{avg}}(\lambda)$) further verifies that by encoding material properties in a single parameter $n_{\textrm{eff}}(\lambda,\theta)$, the model can capture quantitative features in the transmission spectra of integrated LNOI devices.

\subsection{Interferometric resonator}\label{section:IR}
\begin{figure}[!ht]
     \centering
     \begin{subfigure}[b]{0.13\textwidth}
         \centering
         \includegraphics[width=0.9\textwidth]{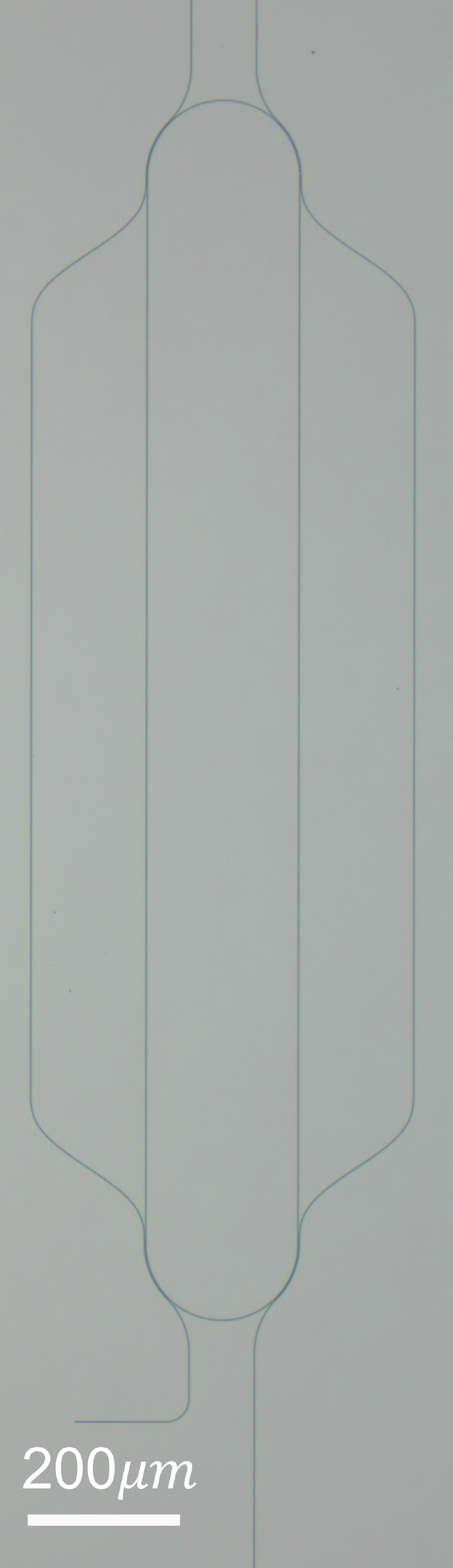}
         \caption{}
         \label{fig:IR_optical}
     \end{subfigure}
     \hfill
     \begin{subfigure}[b]{0.13\textwidth}
         \centering
         \includegraphics[width=\textwidth]{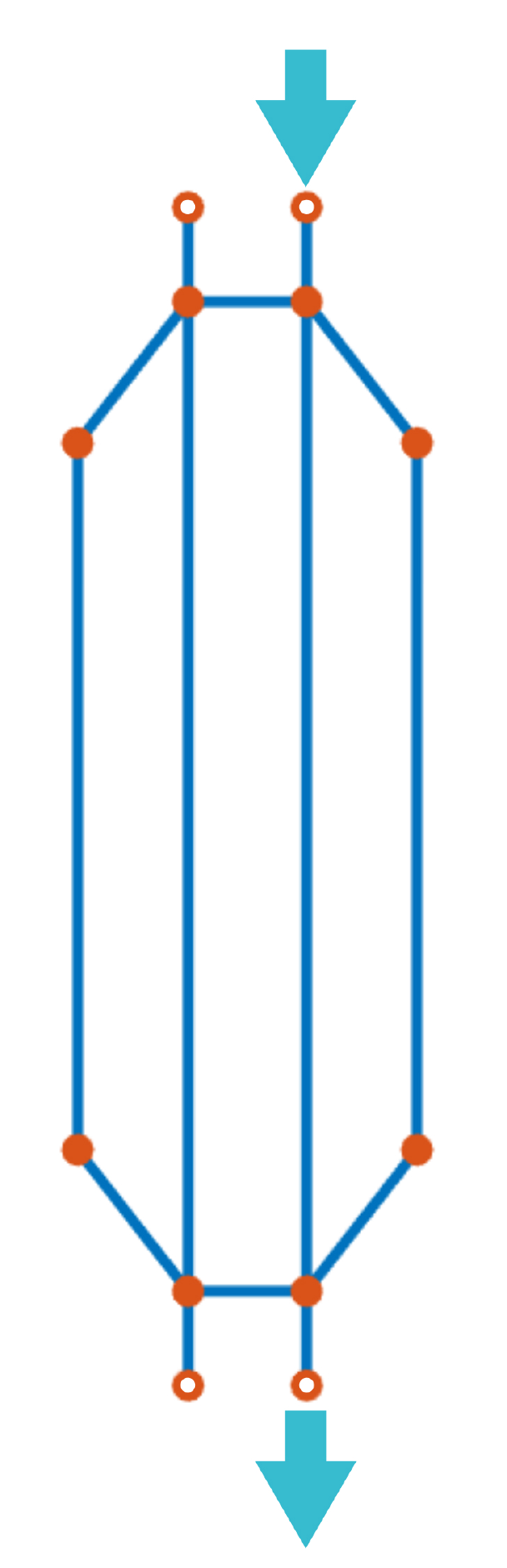}
         
         \caption{}
         \label{fig:IR_scheme}
     \end{subfigure}
     \hfill
     \begin{subfigure}[b]{0.63\textwidth}
         \centering
         \includegraphics[width=\textwidth]{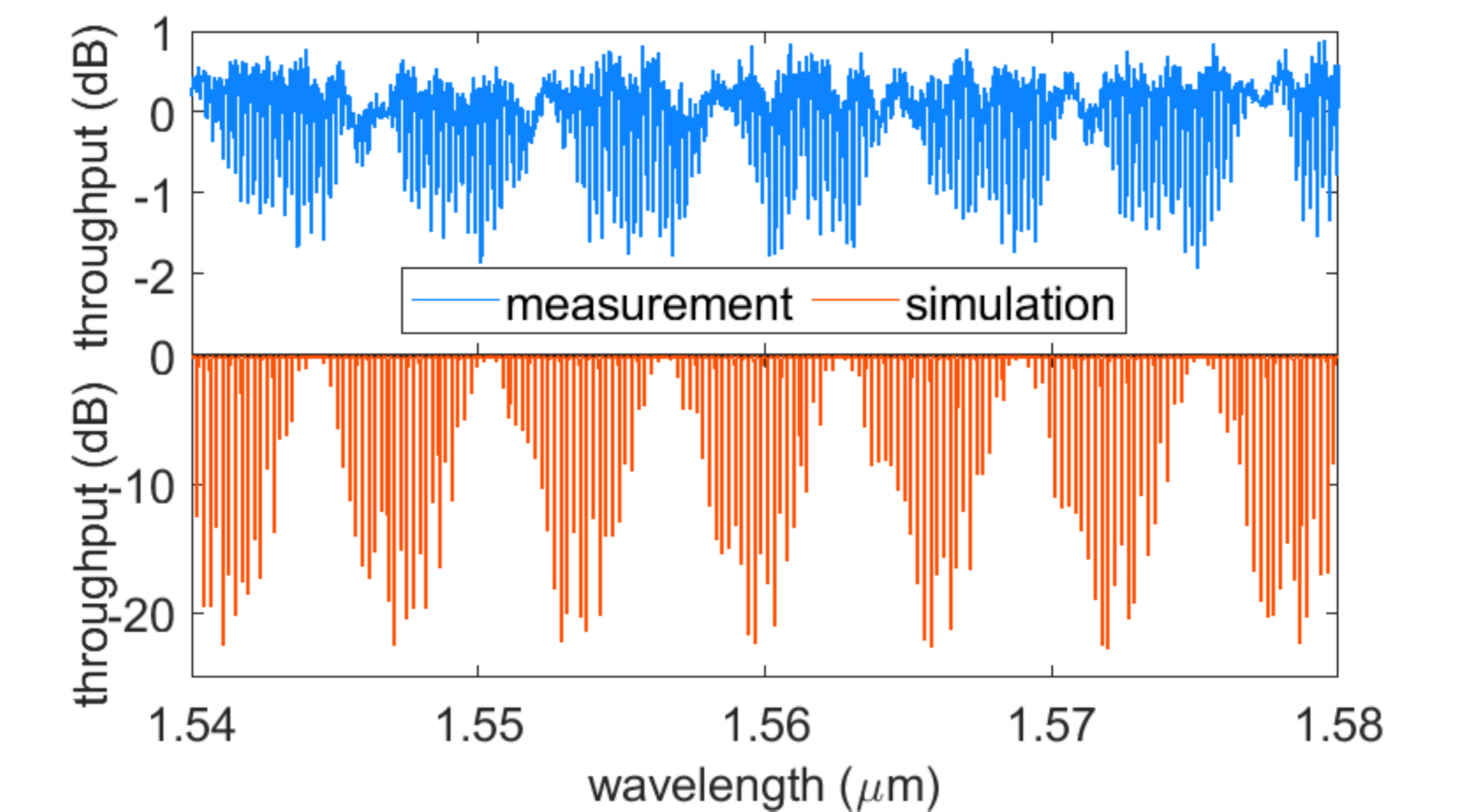}
         \caption{}
         \label{fig:IR_spec}
     \end{subfigure}
        \caption{ Experimental verification on an interferometric resonator. (a) Optical microscope image of an interferometric resonator device. (b) Network geometry used to simulate the design, the arrow indicating the input and output ports for throughput transmission. (c) Throughput port transmission spectrum of the device, measurement (top) vs. simulation (bottom). The measurement data is normalised by fitting and subtracting a second order polynomial, to remove the background spectrum of the superluminescentlight emitting diode source.}
        \label{fig:IR}
\end{figure}

An interferometric resonator is composed of a racetrack-shaped loop and two bus waveguides, each coupling to the loop at two points (Fig.~\ref{fig:IR_optical}). It serves as an example device with more complexity than the single ring resonator \cite{InterferometricResonator_Chen07}, and where different orientations of waveguide components become important. The geometry is approximated by the network schematics shown in Fig.~\ref{fig:IR_scheme}, where the bent sections are approximated by straight waveguides of the same length, and the $n_{\textrm{eff}}(k)$ along the edge is taken to be the same as a straight waveguide placed at the average angle with respect to the thin film. For these devices in the given wavelength range, we have $n_{\textrm{eff}}(\lambda,0)=0.0327\lambda^2 -0.3953 \lambda + 2.4392$, $n_{\textrm{eff}}(\lambda,\pi/2)=-0.0224\lambda^2 -0.2353 \lambda + 2.3687$, and interpolated for the tilted edges $n_{\textrm{eff}}(\lambda,0.6622)=0.0127\lambda^2 -0.3375 \lambda + 2.4141$, with $\lambda$ in $\mu m$ units.

The simulated and measured throughput spectra for such a device are shown in Fig.~\ref{fig:IR_spec}. The simulated spectrum (top) replicates the key features in the measured one, namely the higher frequency resonances of the central loop, as well as the modulating envelop resulting from the unbalanced Mach-Zehnder interferometers formed by the edge of the racetrack and the bus waveguides. A fast Fourier transform on the simulation and measurement data yield a peak at $0.175~nm^{-1}$ and $0.198~nm^{-1}$ respectively, corresponding to 7 and 7.92 envelop cycles in the 40nm window shown in Fig.~\ref{fig:IR_spec}. The discrepancy may be attributed to the measured spectra being less evenly sampled and with more noise present, as well as to the approximation of bent sections by straight edges. Nonetheless, this example demonstrates that even with realistic simplifications to the device geometry, the simulation captures key characteristics in the transmission spectrum of a physical device.

\subsection{Agreement with analytical solutions}
\begin{figure}[!h]
     \centering
     \begin{subfigure}[b]{0.3\textwidth}
         \centering
         \raisebox{0.05\height}{\includegraphics[width=\textwidth]{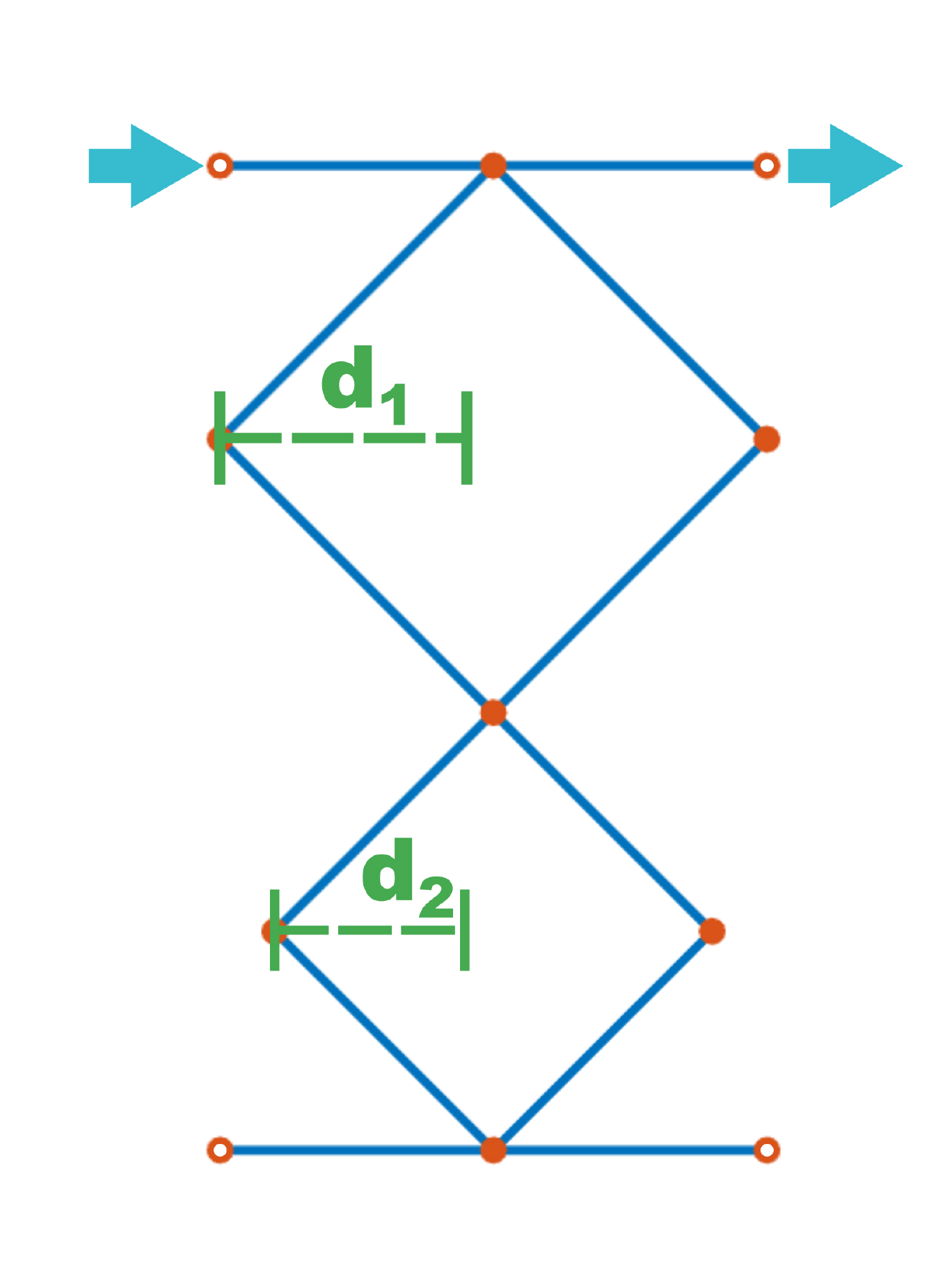}}
         \caption{}
         \label{fig:2rings_scheme}
     \end{subfigure}
     \hfill
     \begin{subfigure}[b]{0.68\textwidth}
         \centering
         \includegraphics[width=\textwidth]{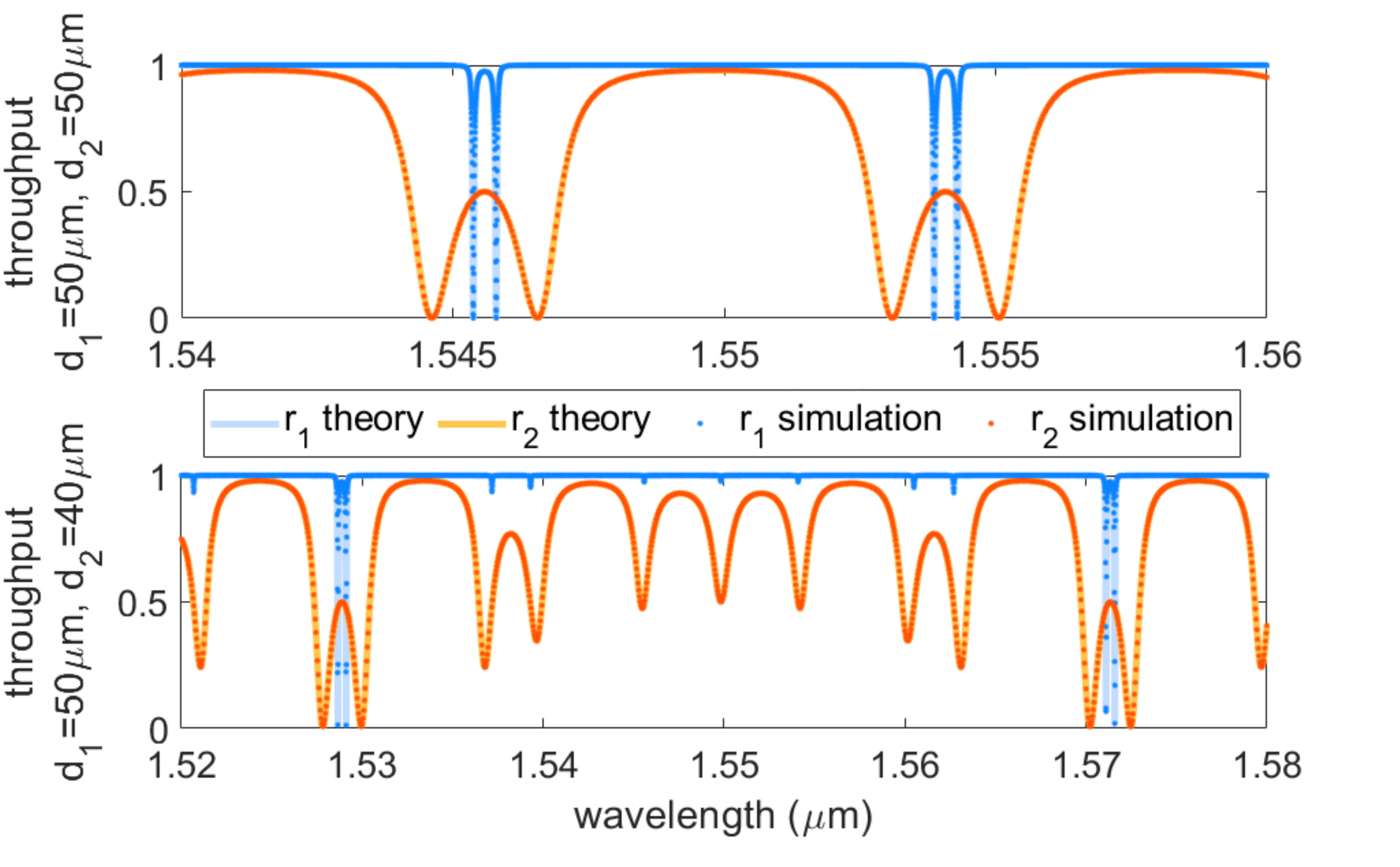}
         \caption{}
         \label{fig:2rings_compare}
     \end{subfigure}
        \caption{ Theoretical verification on serially coupled 2-ring resonators. (a) Graph representation of the device in the model. The throughput configuration, indicated by arrows for input and output ports, is used for computing the spectra. For a lossless and reflection-free system, this captures all the nontrivial transmission information. (b) Transmission spectra of two rings in series, with different ring sizes and coupling strengths at the degree-4 nodes. $r$ is the evanescent coupling parameter as in $s_4$ defined for the ring resonator at the beginning of section \ref{section:RR}, where $r_1=\textrm{cos}(20)$ and $r_2=\textrm{cos}(5)$. Theoretical results follow the analytical solutions given in \cite{RR_2ring}.}
        \label{fig:2rings}
\end{figure}

Besides existing physical samples, there are also photonic networks one may wish to characterise before fabrication, but do not have simple solutions like the single ring resonator. A small-scale network with increased complexity is given by a system of N serially coupled ring resonators. 
We consider the case of N=2 coupled resonators and compare them with the available analytical solution \cite{RR_2ring}, as shown in Fig.~\ref{fig:2rings}. The transmitted spectral shape varies significantly for different ring sizes ($d_1$, $d_2$), as well as different $r$ and $t$ parameters at the degree-4 coupling nodes, all of which are captured exactly by simulation. This demonstrates that for predicting the transmission properties of composite structures, the model performs just as well as traditional analytical methods, without the need for extensive algebraic calculations for each geometry.

Together, the three examples in this section show the scattering matrix model to be a reliable predictor for the transport properties in an integrated LNOI network, while being a drastic simplification from numerical methods such as FDTD and FEM. The combined accuracy and computational tractability thus allow us to apply the scattering matrix model to scaled up optical networks, in particular as a discretised multiple-scattering system.

\section{Developing the model for larger networks} \label{sec:largeGraphs}
Having verified the scattering matrix model for small-scale systems, we consider larger networks characterised by random disorder, which are used as planar, on-chip platforms for multiple scattering of photons. Differently from light transport in 2D random scattering media \cite{2DphotonManagement,Cao2022review}, in random photonic networks scattering and propagation are totally decoupled. Moreover, output channels are limited to few open edges and no further spatial discretisation is needed.  When all these features are utilised,  advanced control on light transport and deposition throughout the network can be achieved. 
A key feature of this model is the ability to introduce arbitrary scattering functions at the nodes of a photonic network. We present numerical experiments that demonstrate how tuning the scattering anisotropy at individual nodes modifies overall transport properties. This is a useful tool for many applications requiring control  of the dwell time of light within the network without affecting the network's size .

\subsection{Randomised grid-like networks}
\begin{figure}[!ht]
     \centering
     \begin{subfigure}[c]{0.5\textwidth}
         \centering
         \includegraphics[width=\textwidth]{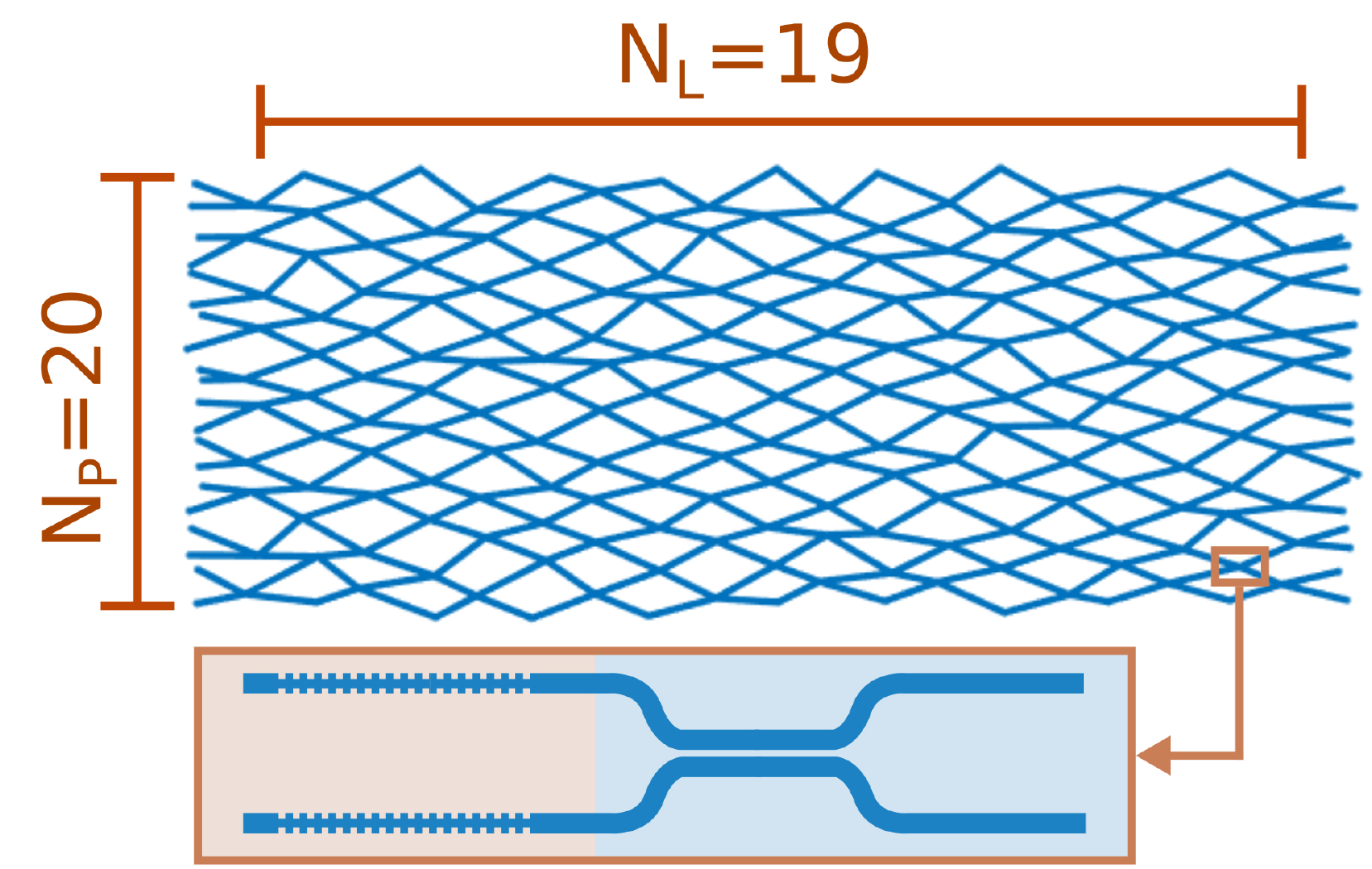}
         \caption{}
         \label{fig:Uscheme}
     \end{subfigure}
     \hfill
     \begin{subfigure}[c]{0.45\textwidth}
         \centering
         \includegraphics[width=\textwidth]{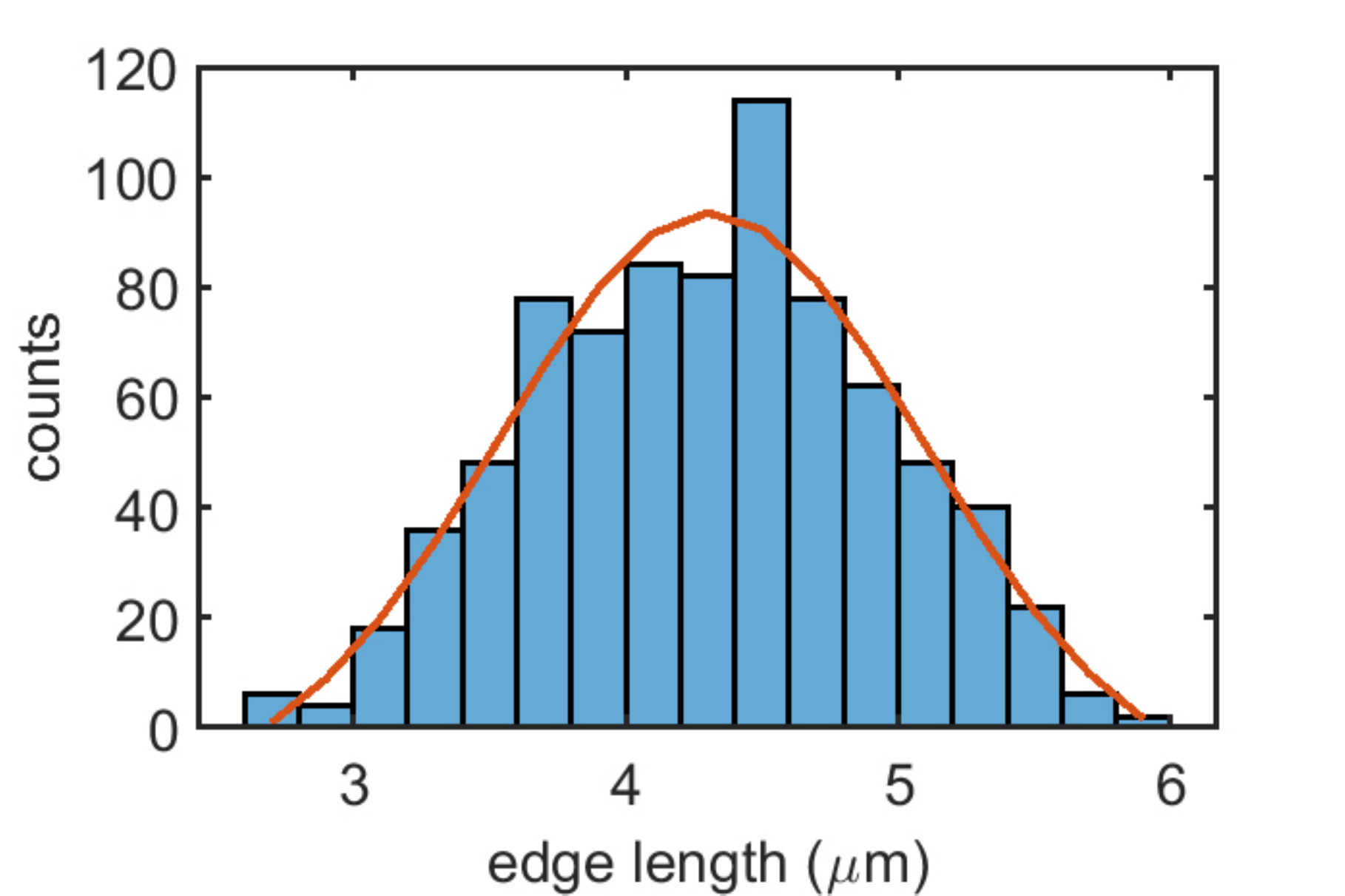}
         \caption{}
         \label{fig:Uedgelength}
     \end{subfigure}

        \caption{(a) A randomised grid-like network (node markers omitted for legibility): light can enter and exit through the open ports on left and right, propagate along waveguides denoted by the lines, and scatter at nodes where the lines meet. The size of the network is defined by $N_P$, the number of open ports on one side of the network, and can be thought of as the width, and $N_L$, the number of internal layers. Inset: the node design corresponding to the chosen scattering matrix (Eq.~\ref{eq:abmatrix}) consists of a beamsplitter section (blue underlay) and Bragg reflector section (orange underlay). (b) Edge length distribution of the network on the left, fitted with a Gaussian distribution. The average vertical distance between nodes is $3\mu m$, horizontal $4\mu m$, random displacement of each node uniformly sampled from $[0,1.5]\mu m$ in each direction. For wavelengths around $1500nm$, this ensures a fully randomised phase between two scattering events.}
        \label{fig:Uinterferometer}
\end{figure}

The geometry that we select as a model to investigate light scattering in networks is inspired by unidirectional forward-scattering optical circuits, which include beam splitters (realised via evanescent couplers) and phase shifters to couple and manipulate light modes travelling in waveguides: for example, to achieve universal unitary gates in quantum optical circuits \cite{UInterferometer}, and in particular LNOI quantum photonic processors \cite{LNOIquantumProcessor}. This is translated into grid-like networks with degree-4 nodes which, under the appropriate scattering matrix, can act exactly as forward-scattering beam splitters (Fig.~\ref{fig:Uscheme}). The size of the network is characterised by two parameters: the width, or the number of ports on each side $N_P$, and the depth, or the number of scattering layers $N_L$. Starting from regular grids, we transition into the multiple scattering regime by including backscattering at the nodes, and introduce disorder by adding a random displacement to each node, resulting in Gaussian distribution in the edge lengths (Fig.~\ref{fig:Uedgelength}). This induces a random phase that is picked up by light between two scattering events. The relevant geometric parameters are detailed in the caption of Fig.~\ref{fig:Uinterferometer}.

The scattering matrix at the degree-4 nodes are selected to be of the form
\begin{equation} \label{eq:abmatrix}
s_4 = 
\begin{pmatrix}
\sqrt{\beta} & 0 & \sqrt{(1-\alpha)(1-\beta)} & \sqrt{\alpha(1-\beta)}\\
0 & \sqrt{\beta} & -\sqrt{\alpha(1-\beta)} & \sqrt{(1-\alpha)(1-\beta)} \\
\sqrt{(1-\alpha)(1-\beta)} & -\sqrt{\alpha(1-\beta)} & -\sqrt{\beta} & 0 \\
\sqrt{\alpha(1-\beta)} & \sqrt{(1-\alpha)(1-\beta)} & 0 & -\sqrt{\beta} 
\end{pmatrix}
\end{equation}
with $\alpha,\beta\in[0,1]$ being the splitting and backreflection parameters, respectively. $s_4$ is unitary and symmetric, and for $\beta=0$ reduces to a unidirectional forward-scattering beam splitter. Such matrices are not fully isotropic due to the presence of zero entries; they are nevertheless selected for the ease of parametrisation while satisfying the constraints of an ideal physical device (lossless and reciprocal). Such a matrix is realised experimentally by attaching two Bragg reflectors with reflectivity $\beta$ to a beam splitter with splitting ratio $\alpha$ (see inset of Fig.~\ref{fig:Uscheme}). 

Due to the random nature of the network, the transmission at the output edges creates discretised speckle-like patterns which we use to monitor transport properties. The model can simulate solutions independently at a range of different wavelengths, which together generate transmission spectra between arbitrary input and output ports.

\subsection{Transmission and internal field}
Given the geometry and scattering behaviour described above, a network is specified by parameters $N_P$, $N_L$, $\alpha$ and $\beta$. In the following studies, we keep the network width $N_P=30$ and node splitting ratio $\alpha=0.5$ fixed, and vary the network depth $N_L$ and node reflectivity $\beta$.

\begin{figure}[!h]
\centering
\begin{subfigure}[b]{0.32\textwidth}
    \includegraphics[width=\textwidth]{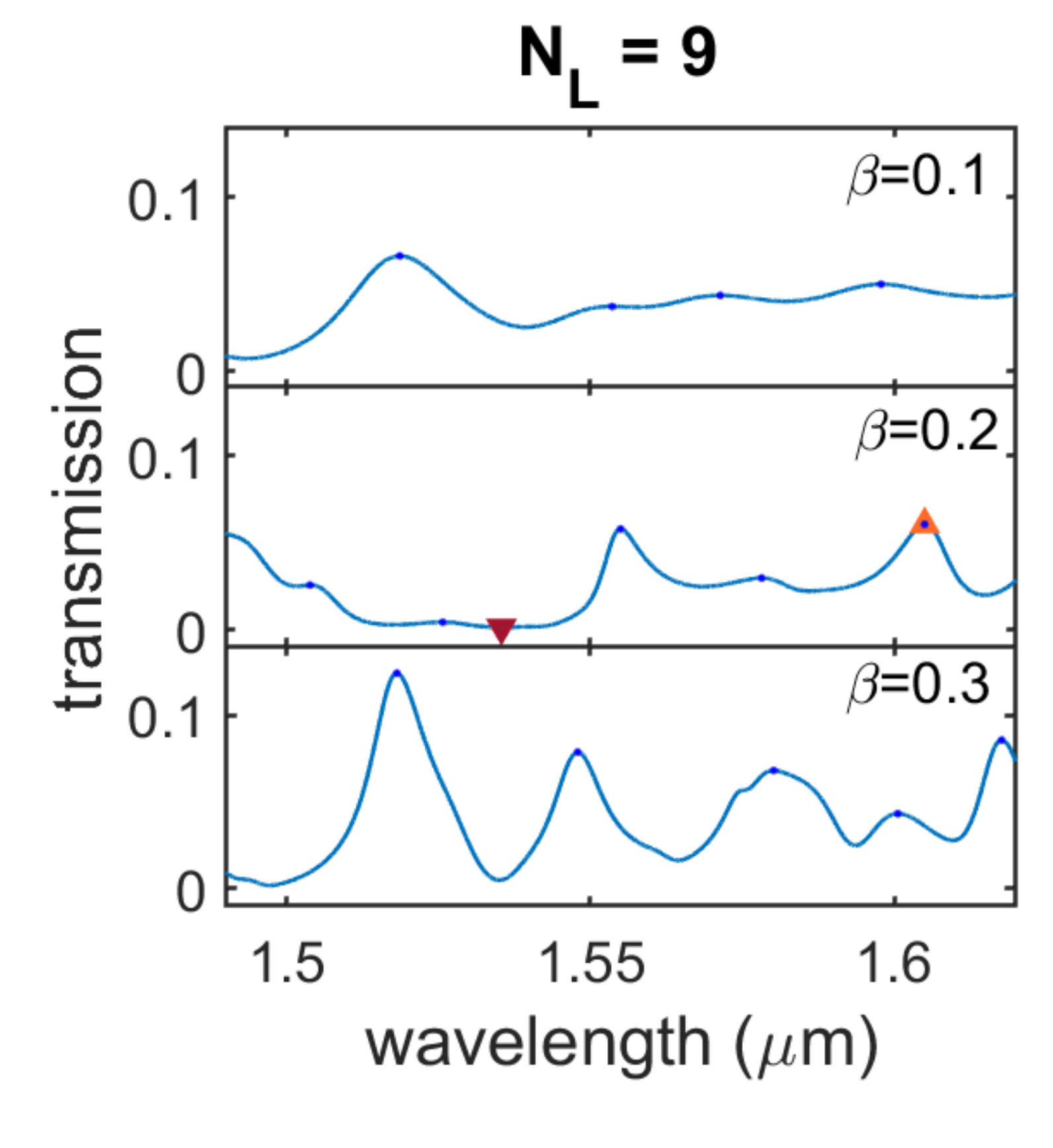}
    \vspace*{-30pt}
    \subcaption{}
    \vspace*{30pt}
    \label{fig:30x9spec}
\end{subfigure}%
\begin{subfigure}[b]{0.32\textwidth}
    \includegraphics[width=\textwidth]{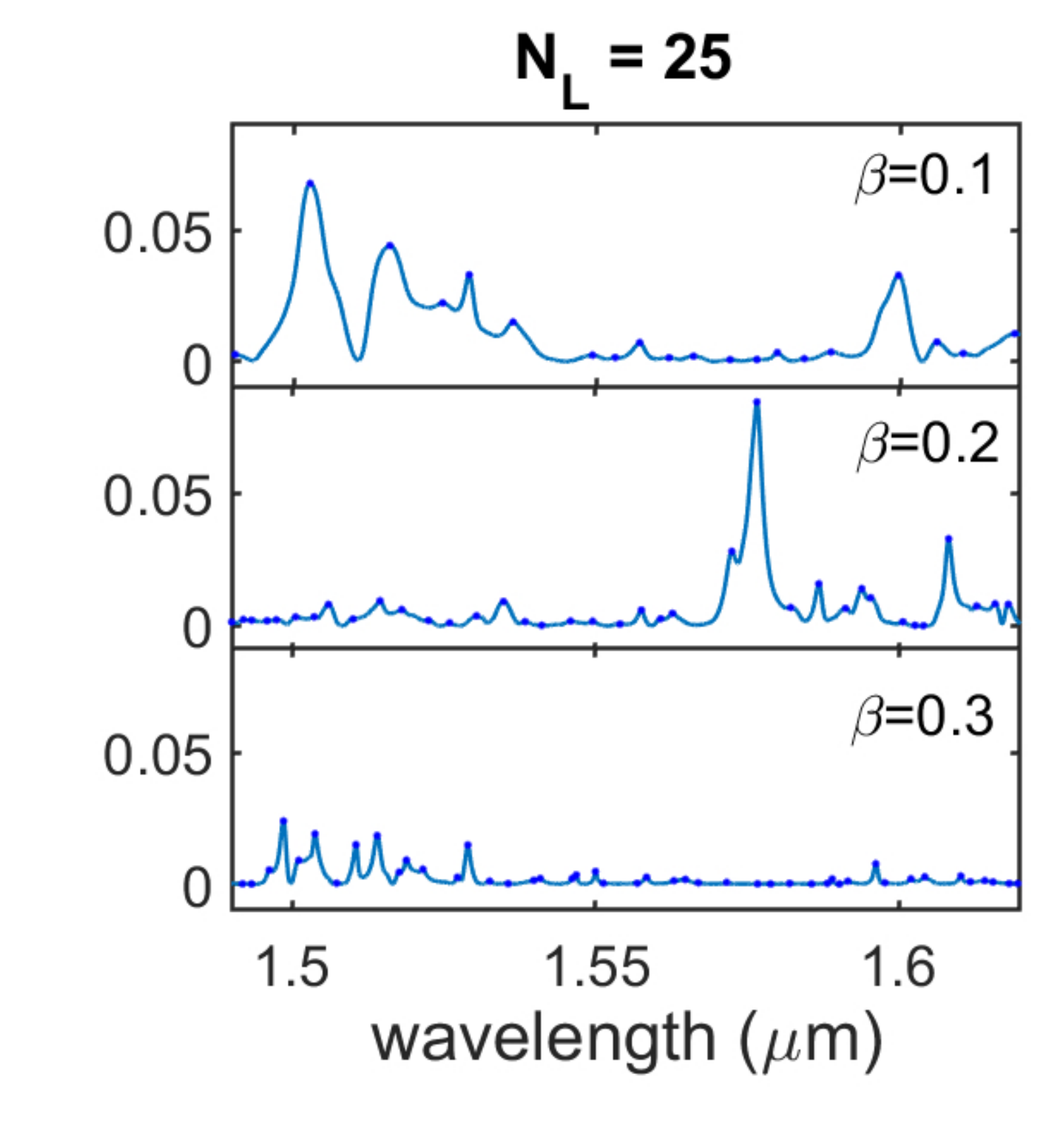}
    \vspace*{-30pt}
    \subcaption{}
    \vspace*{30pt}
    \label{fig:30x25spec}
\end{subfigure}%
\begin{subfigure}[b]{0.32\textwidth}
\includegraphics[width=\textwidth]{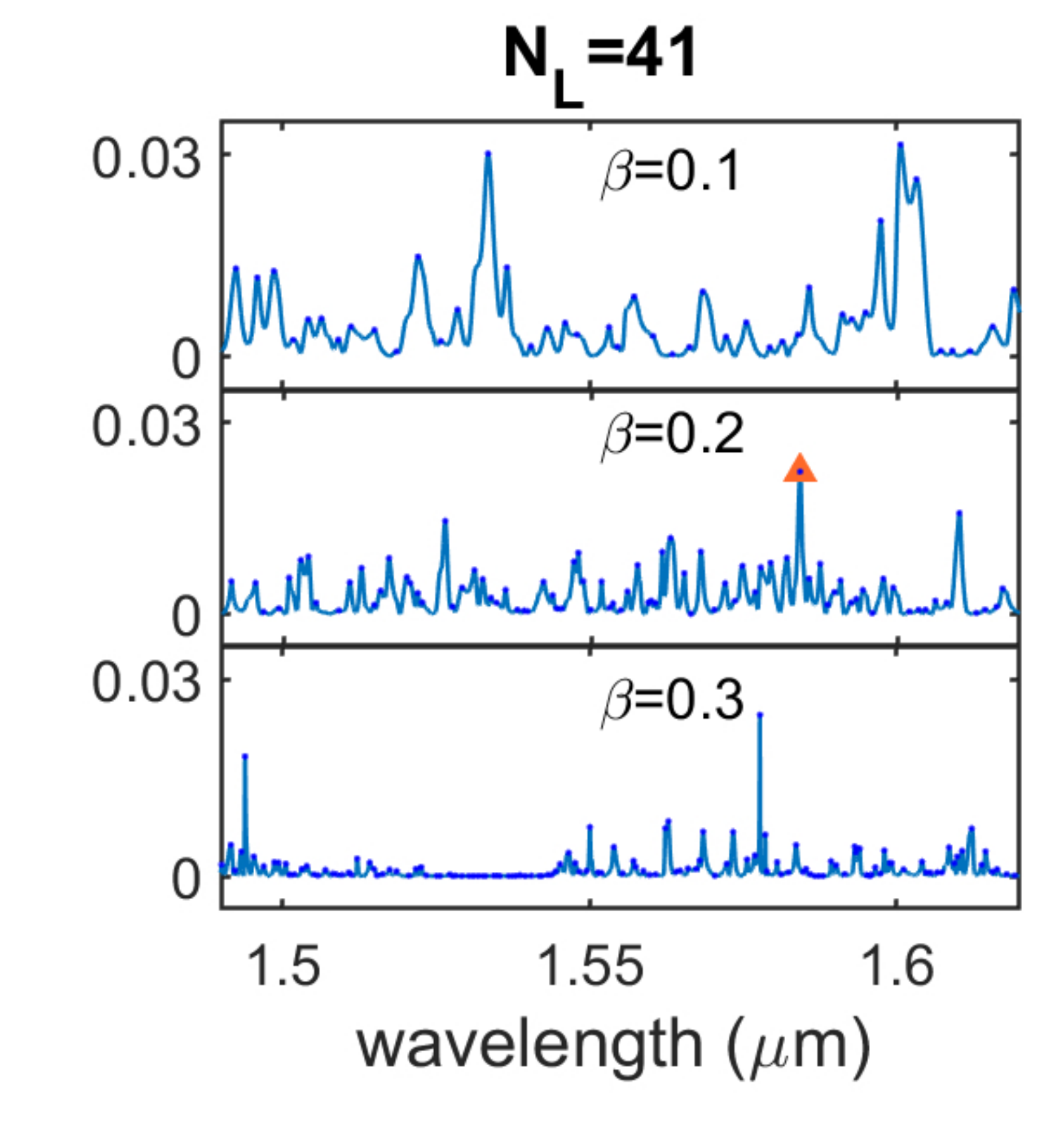}
    \vspace*{-30pt}
    \subcaption{}
    \vspace*{30pt}
\label{fig:30x41spec}
\end{subfigure}
    \vspace*{-25pt}
\caption{Transmission spectra between ports 10 and 40, for networks with $N_P=30$ and (a) $N_L=9$, (b) $N_L=25$, and (c) $N_L=41$. From top to bottom each geometry had local scattering matrices with reflectivities $\beta=0.1$,0.2,0.3, respectively. As $N_L$ increases, the peaks become narrower, similar to the transmission peak sizes in other multiple scattering systems. Furthermore, we see an increase in the number of peaks as $\beta$ is increased. The triangles mark the maximum and minimum transmission points, for which the corresponding field distribution are shown in Fig.~\ref{fig:network_fields}.}
\label{fig:network_spectra}
\end{figure}

For each network, the transmission spectrum between any two open ports can be computed by selecting the relevant element in the overall scattering matrix $S_{io}$. This is a much larger set of possible measurements compared to smaller devices described in the previous section. A few example spectra of varying network sizes $N_L=\{ 9,25,41\}$ and backscattering strengths $\beta=\{ 0.1,0.2,0.3\}$ are shown in Fig.~\ref{fig:network_spectra}. Increasing $N_L$ and $\beta$ produces sharper resonances, as well as more complex spectra with more resonance peaks.  

\begin{figure}[!h]
\centering
\begin{subfigure}[c]{0.22\textwidth}
    \includegraphics[width=\textwidth]{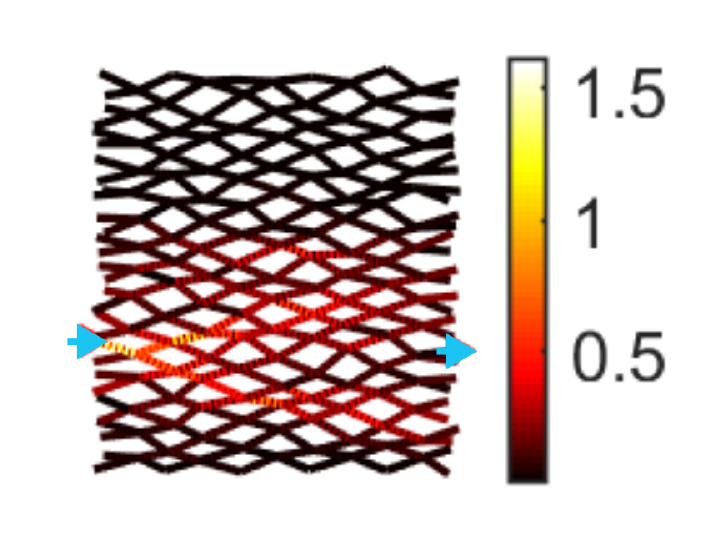}
    \caption{}
    \label{fig:30x9max}
\end{subfigure}%
\begin{subfigure}[c]{0.22\textwidth}
    \includegraphics[width=\textwidth]{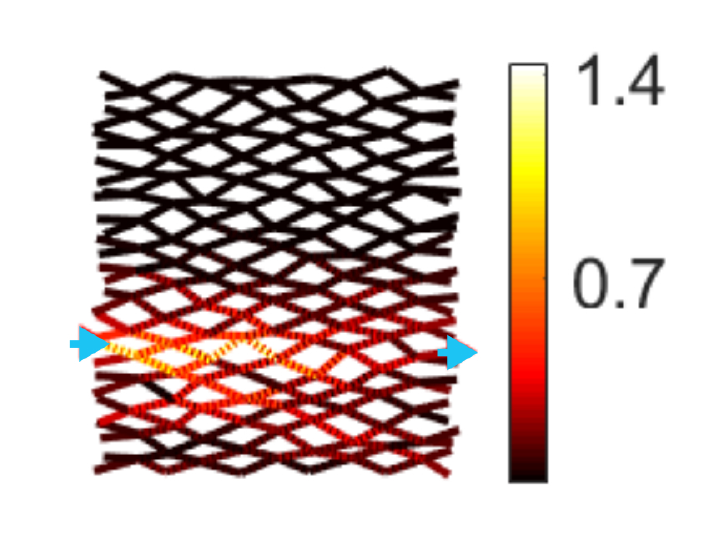}
    \caption{}
    \label{fig:30x9min}
\end{subfigure}%
\begin{subfigure}[c]{0.56\textwidth}
\vspace*{5pt}
\includegraphics[width=\textwidth]{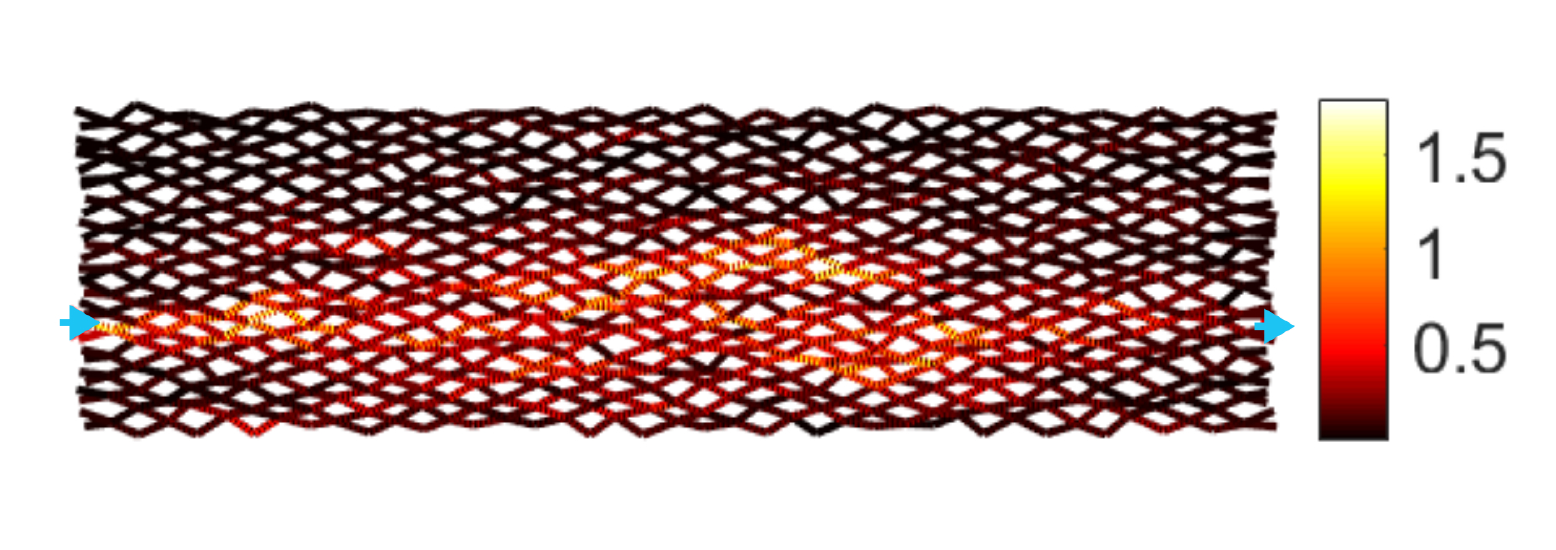}
\vspace*{-20pt}
\caption{}
\vspace*{10pt}
\label{fig:30x41max}
\end{subfigure}
\caption{Internal field distributions for $N_P=30$, $\beta=0.2$ and (a) (b) $N_L=9$, or (c) $N_L=41$. Ports 10 and 40 are labelled with blue markers, and unit-power input is injected at port 10. (a) and (b) show the minimum and maximum transmission wavelengths respectively, whereas (c) shows the maximum transmission for a larger network.}
\label{fig:network_fields}
\end{figure}

Since the graph model contains solutions on the entire network, one can go beyond input-output relations and look at internal field distributions. This information is inaccessible in experiment without introducing additional out of plane losses, and the simulation therefore provides additional insight into the scattering process for different network geometries. Fig.~\ref{fig:network_fields} shows the corresponding field distribution at maximum and minimum transmission wavelengths for a subset of the spectra shown in Fig.~\ref{fig:network_spectra}. For the $N_L=9$ network, the field profiles show a distinct difference between high and low transmission between specified input and output ports: in the minimum case (Fig.~\ref{fig:30x9min}), the field is split and sent to either side of the target output, whereas for maximum transmission (Fig.~\ref{fig:30x9max}) it is focused on the same port. For the larger network (Fig.~\ref{fig:30x41max}), high transmission is generated by exciting a series of coupled resonances between the input and output, akin to the necklace states observed in other multiple scattering systems \cite{1DNecklaceStates}. The excited field has a higher intensity near the centre of the network, in contrast to the linear decay with network depth in the diffusive regime.

\begin{figure}[!ht]
     \begin{minipage}{0.25\textwidth}%
        \begin{subfigure}[b]{\textwidth}
         \centering
         \includegraphics[width=\textwidth]{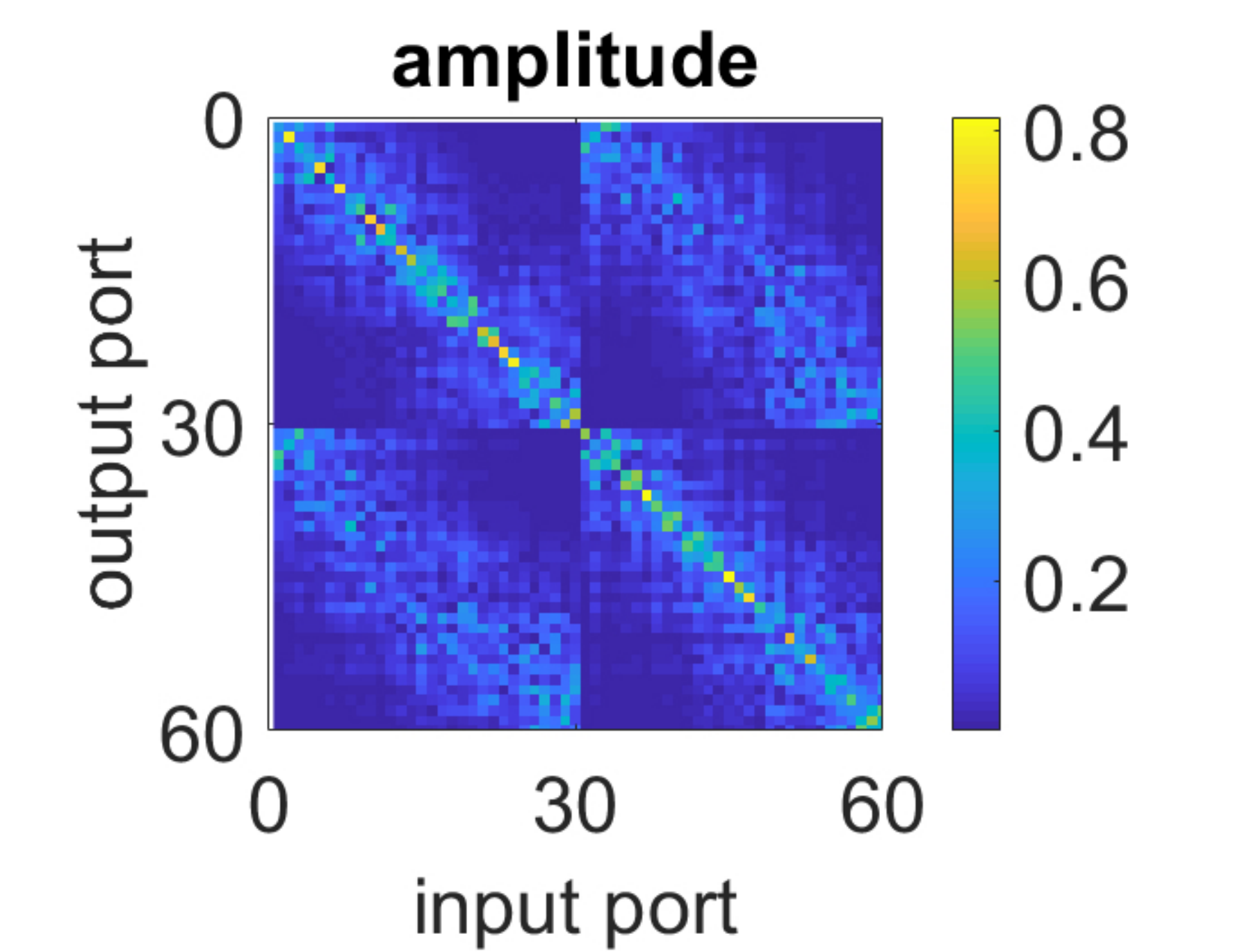}
         \caption{}
         \label{fig:T_intensity}
     \end{subfigure}
     \hfill
     \begin{subfigure}[b]{\textwidth}
         \centering
         \includegraphics[width=\textwidth]{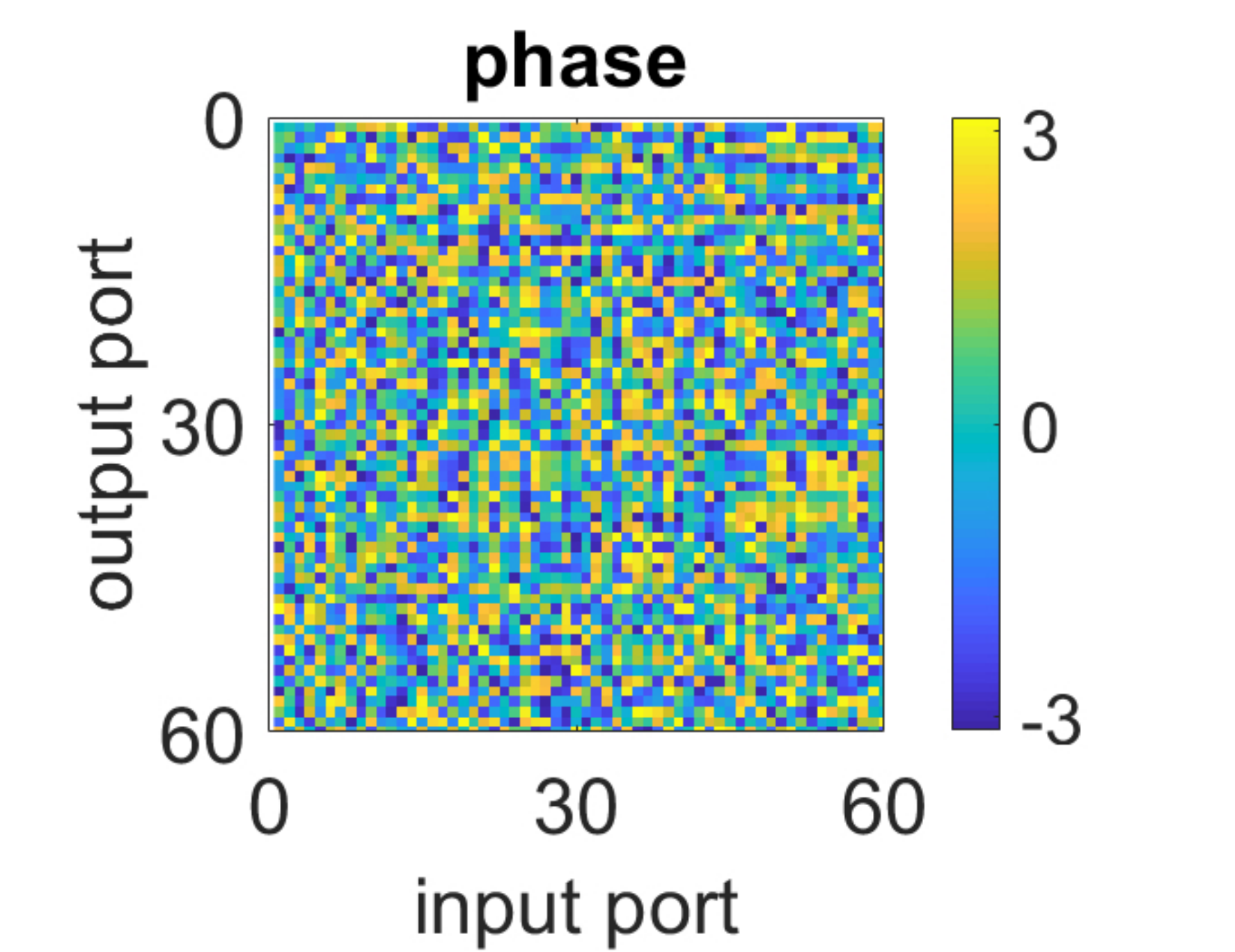}
         \caption{}
         \label{fig:T_phase}
     \end{subfigure}
    \end{minipage}%
    \begin{minipage}{0.40\textwidth}
     \begin{subfigure}[b]{\textwidth}
         \centering
         \includegraphics[width=\textwidth]{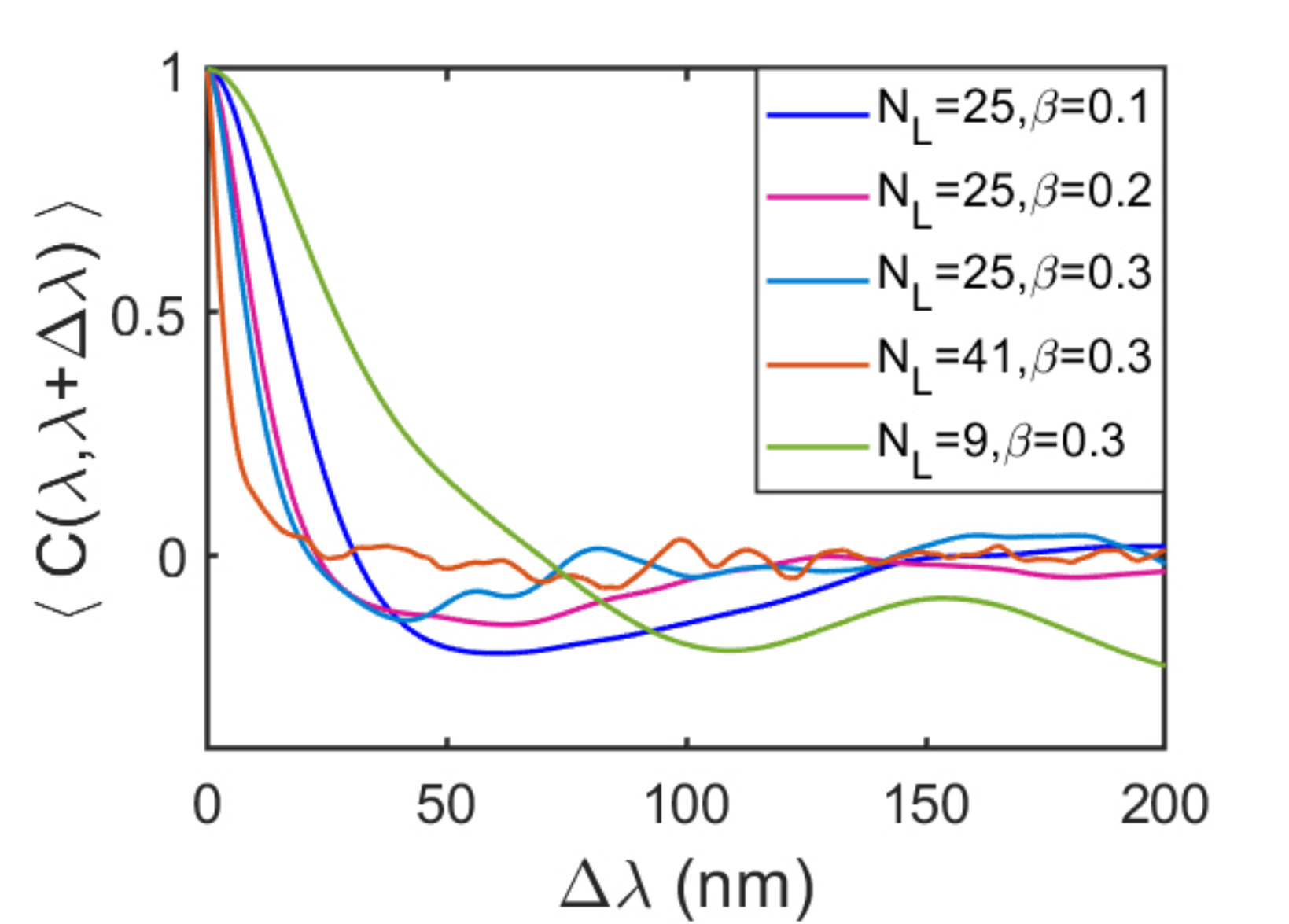}
         \vspace*{-25pt}
         \caption{}
         \vspace*{25pt}

         \label{fig:speccorr}
     \end{subfigure}
     \end{minipage}%
     \begin{minipage}{0.35\textwidth}
     \begin{subfigure}[b]{\textwidth}
         \centering
         \includegraphics[width=\textwidth]{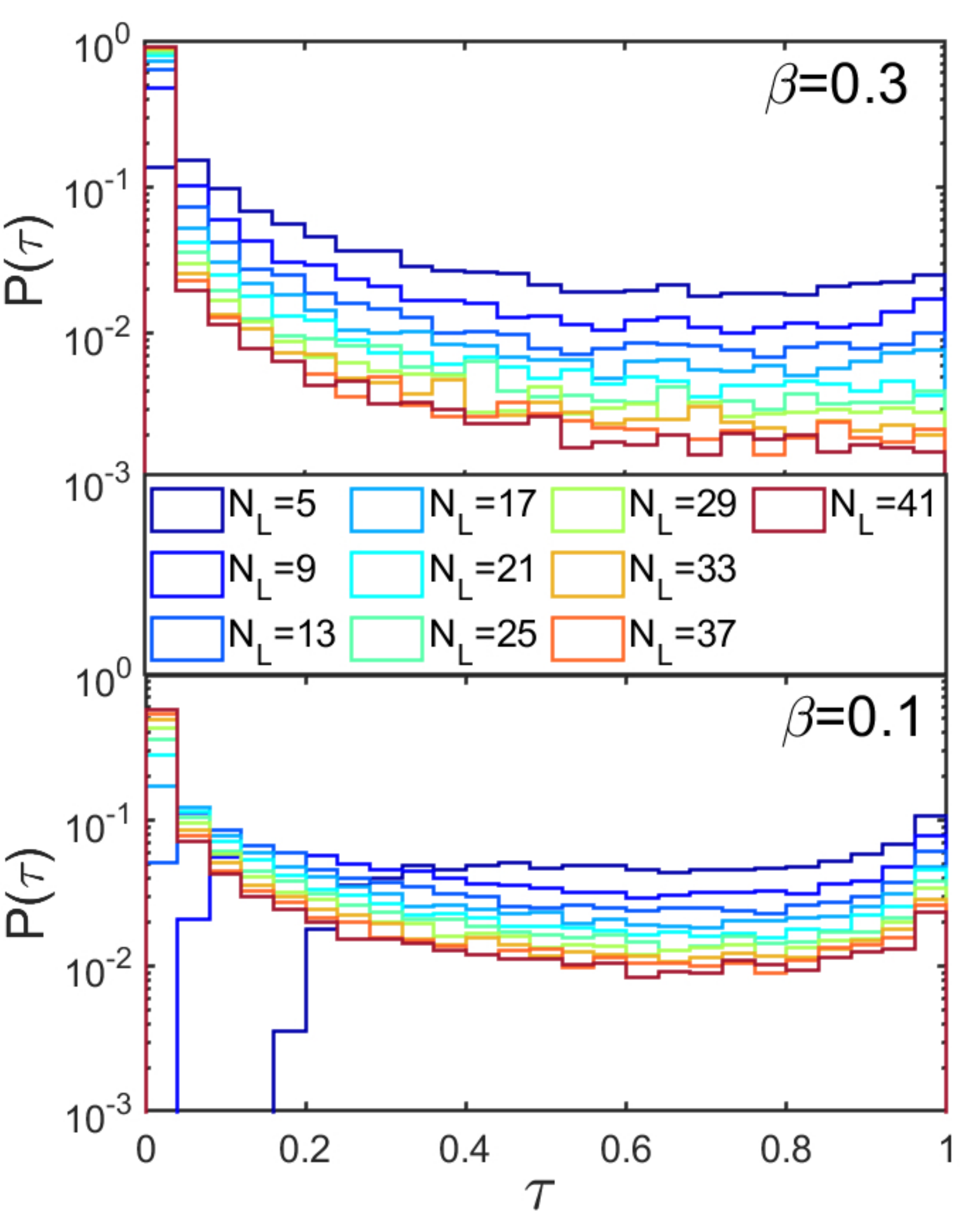}
         \vspace*{-20pt}
         \caption{}
         \vspace*{20pt}
         \label{fig:T_eigenval_beta}
     \end{subfigure}
     \end{minipage}
    
        \vspace*{-20pt}
        \caption{The $S_{io}$ matrix and transmission behaviour. (a) Amplitude and (b) phase components of the $S_{io}$ matrix. Ports 1-30 are on the left side of the network, 31-60 on the right. Each column is the speckle pattern resulting from a single-port input at a single wavelength. (c) Spectral correlation between speckle patterns, averaged over all inputs and 500 random realisations. Sharper spectral resolution is observed for larger $N_L$ or higher $\beta$. (d) Transmission eigenspectra for networks with $\beta=0.1$ and $0.3$, with varying depths, averaged over 500 random realisations. The increase in the distribution at higher transmittance is observed, but the effect is diminished for higher $\beta$.}
        \label{fig:speckle_and_eigen}
\end{figure}

The individual transmission channels provide a convenient experimental analogue to smaller devices, but for large networks it is instructive to look at the entire scattering matrix $S_{io}$, an example of which is shown in Fig.~\ref{fig:T_intensity} (amplitude) and \ref{fig:T_phase} (phase). Since the open ports are divided into the ones on the left and the right of a network (Fig.~\ref{fig:Uscheme}), the $S_{io}$ matrix is divided into four quadrants: two diagonal blocks of \textit{reflection} back to the same side, and two off-diagonal blocks of \textit{transmission} to the opposite side. We write this as $S = \begin{pmatrix}R_{LL} & T_{RL} \\ T_{LR} & R_{RR} \end{pmatrix}$, where $T:=ðT_{LR}=T_{RL}^t$ due to reciprocity. The separation between $T$ and $R$ is clearly visible in the amplitude matrix, whereas the phase matrix shows fully mixed phases between $-\pi$ and $\pi$. 
Furthermore, each column in $S_{io}$ is the speckle pattern resulting from a single-port input, with a reflected component (in $R_{LL}$ or $R_{RR}$) and a transmitted component (in $T_{LR}$ or $T_{RL}$). The wavelength-dependent speckle generated by the random geometry is a resource that can be utilised, for example in a random spectrometer \cite{random_spectometer_2013}. The spectral resolution is provided by decorrelation between the speckle patterns at different wavelengths. Fig.~\ref{fig:speccorr} shows the transmitted speckle spectral correlation for networks with different $N_L$ and $\beta$, averaged over all input ports and wavelengths. A higher spectral resolution is observed for networks with larger sizes, as well as stronger backreflections at the nodes. The observed behaviour can be interpreted in the complementary view of delay time in multiple scattering processes:  the resolution of an interferometer is improved  when longer optical paths come into play. Therefore, spectral decorrelation results confirm the capability of this platform to increase light circulation via multiple scattering, and show that average properties can be tuned by the parameters $\beta$ and $N_L$.

One can also go beyond the speckle pattern from single-port inputs, and study the transmission matrix $T$ as a whole by decomposing them into transmission eigenchannels. Fig.~\ref{fig:T_eigenval_beta} shows the eigenvalues of $T^\dagger T$, averaged over 500 random realisations, of networks with $\beta=0.1$ and 0.3, and at varying depths. Here a realisation means one instance of a network generated with the specified type of randomness, i.e. uniform random displacement of the nodes. Ergodicity is the equivalence between averaging over wavelengths and realisations, and the reason why networks satisfy this condition will become clear in section 4.3. The abundance of open channels, i.e. an increase in density of near-unity eigenvalues is observed, in accordance with the bimodal distribution expected for continuous random scattering \cite{RotterGigan2017Review}. This effect is diminished for higher $\beta$, suggesting excessive backreflection at the nodes results in transmission deviating from the diffusion regime. This deviation can also been seen in the scaling of transmission with network size, discussed below. 

\subsection{Transport properties on a discrete length scale}
To statistically study transport properties on network ensembles with different design parameters, we look at the scalar value $T_{total}$, i.e. the total transmitted power from unit input. Scaling laws, which relate $T_{total}$ to system size, are a useful tool for extracting information from multiple scattering media, for example to measure the transport mean free path $l^*$ and the localisation length $\xi$. Here we study both $l^*$ and $\xi$ on random networks, with the added caveat that network size is better characterised by discrete measures ($N_L$) rather than the physical size of the system. If one extends all edges in the network by a certain factor, a different random phase will be traversed between any pair of connected nodes, but the overall nature of the random scattering remains unchanged (see Fig.~\ref{fig:Uedgelength}: the edge length distribution will remain Gaussian, with sufficient variance to accommodate $2\pi$ phase difference between edges). Therefore, stretching or shrinking the size of a network is equivalent to generating another random realisation of a network of the same size. Since stretching the edge lengths is complementary to reducing wavelength, this edge-length-invariance (combined with wavelength-independent scattering at the nodes) further implies that the network scattering matrices are ergodic, meaning it is statistically equivalent to average over wavelengths and random realisations. The number of layers $N_L$  is therefore is chosen to be the discrete measure of size in our grid-like networks, instead of the physical scale. As a result, the computed $\xi$ and $l^*$ only obtain a physical length scale by relating them to relevant dimensions in the network.

\begin{figure}[!ht]
    \begin{subfigure}[b]{0.5\textwidth}
                 \centering
                 \includegraphics[width=0.98\textwidth]{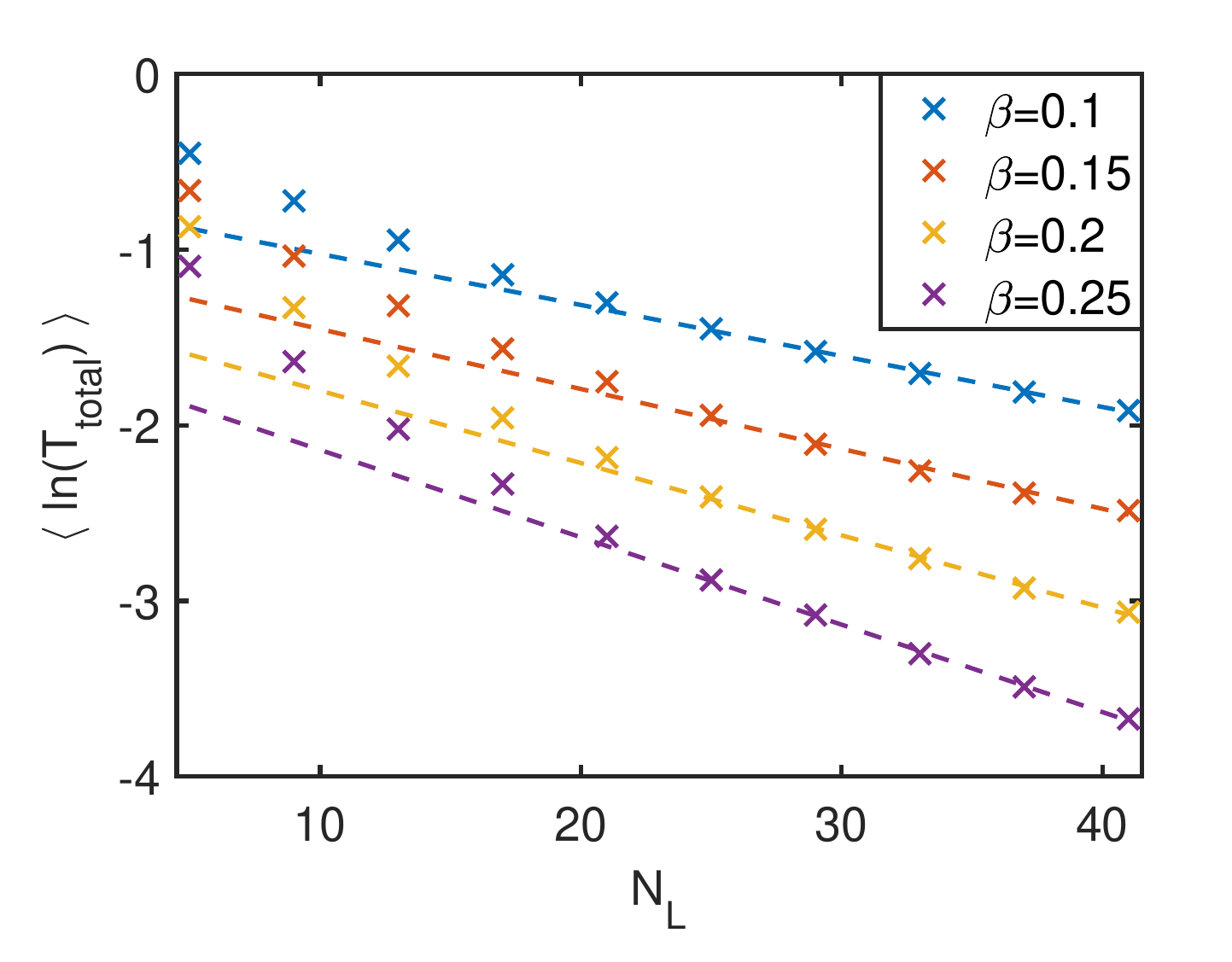}
                 \vspace*{-25pt}
                 \caption{}
                 \vspace*{25pt}
                 \label{fig:lnTScaling}
         \end{subfigure}
    \begin{subfigure}[b]{0.5\textwidth}
                 \centering
                 \includegraphics[width=0.98\textwidth]{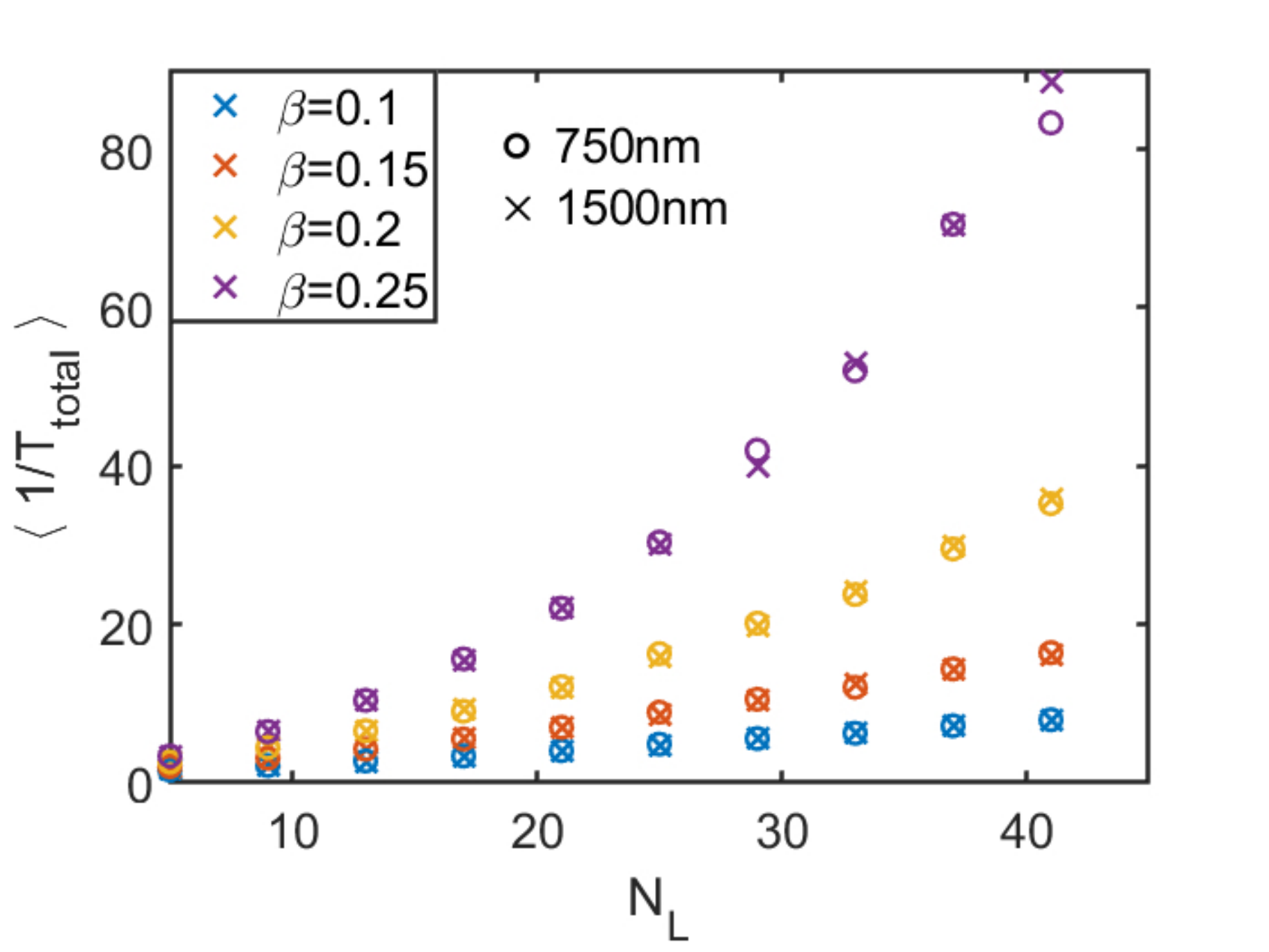}    
                 \vspace*{-25pt}
                 \caption{}
                 \vspace*{25pt}
                 \label{fig:TinvScaling}
        \end{subfigure}%

         \vspace*{-20pt}
    \begin{subfigure}[c]{0.3\textwidth}
                 \centering
                 \includegraphics[width=0.98\textwidth]{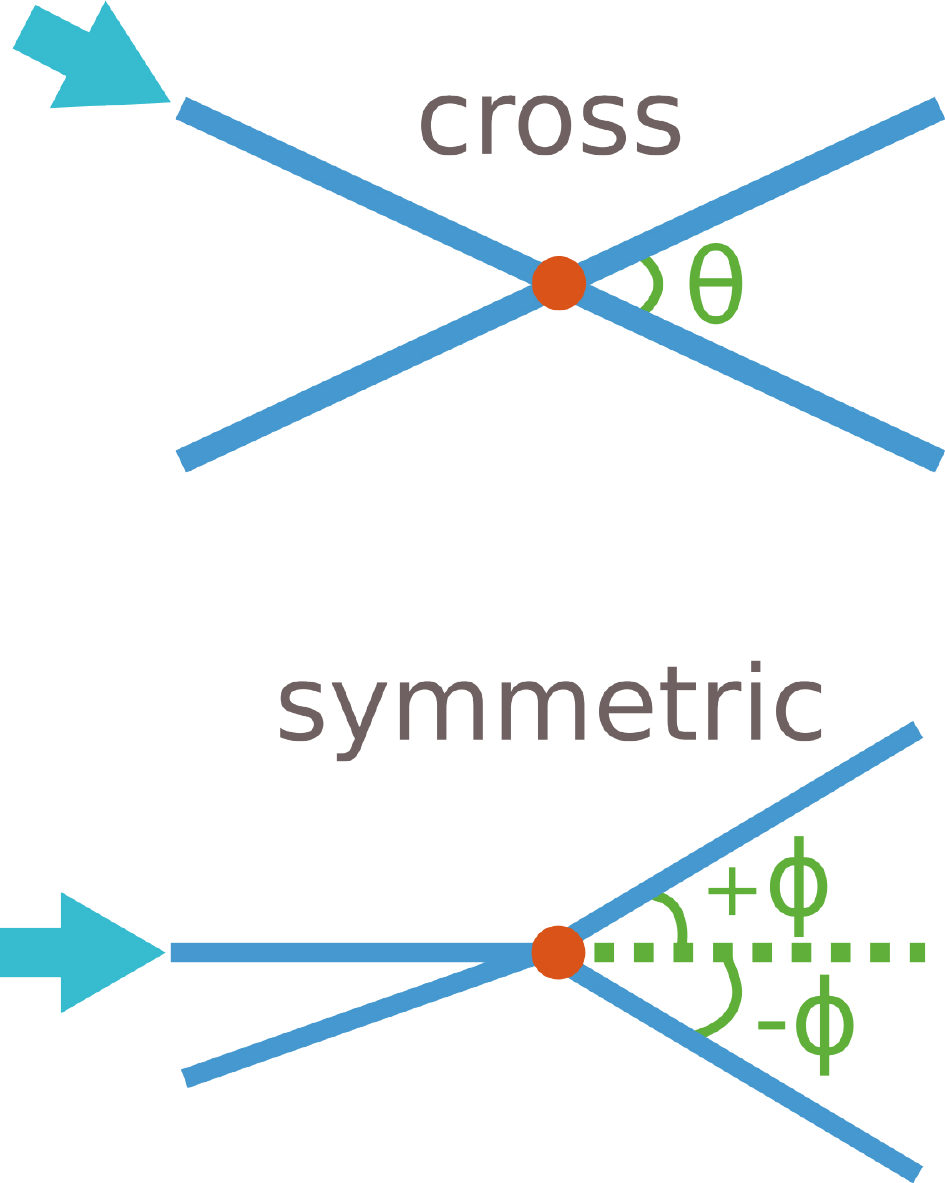}
                 \vspace*{-10pt}
                 \caption{}
                 \vspace*{10pt}
                 \label{fig:node_g_schemes}
         \end{subfigure}
    \begin{subfigure}[c]{0.32\textwidth}
                 \centering
                 \includegraphics[width=0.98\textwidth]{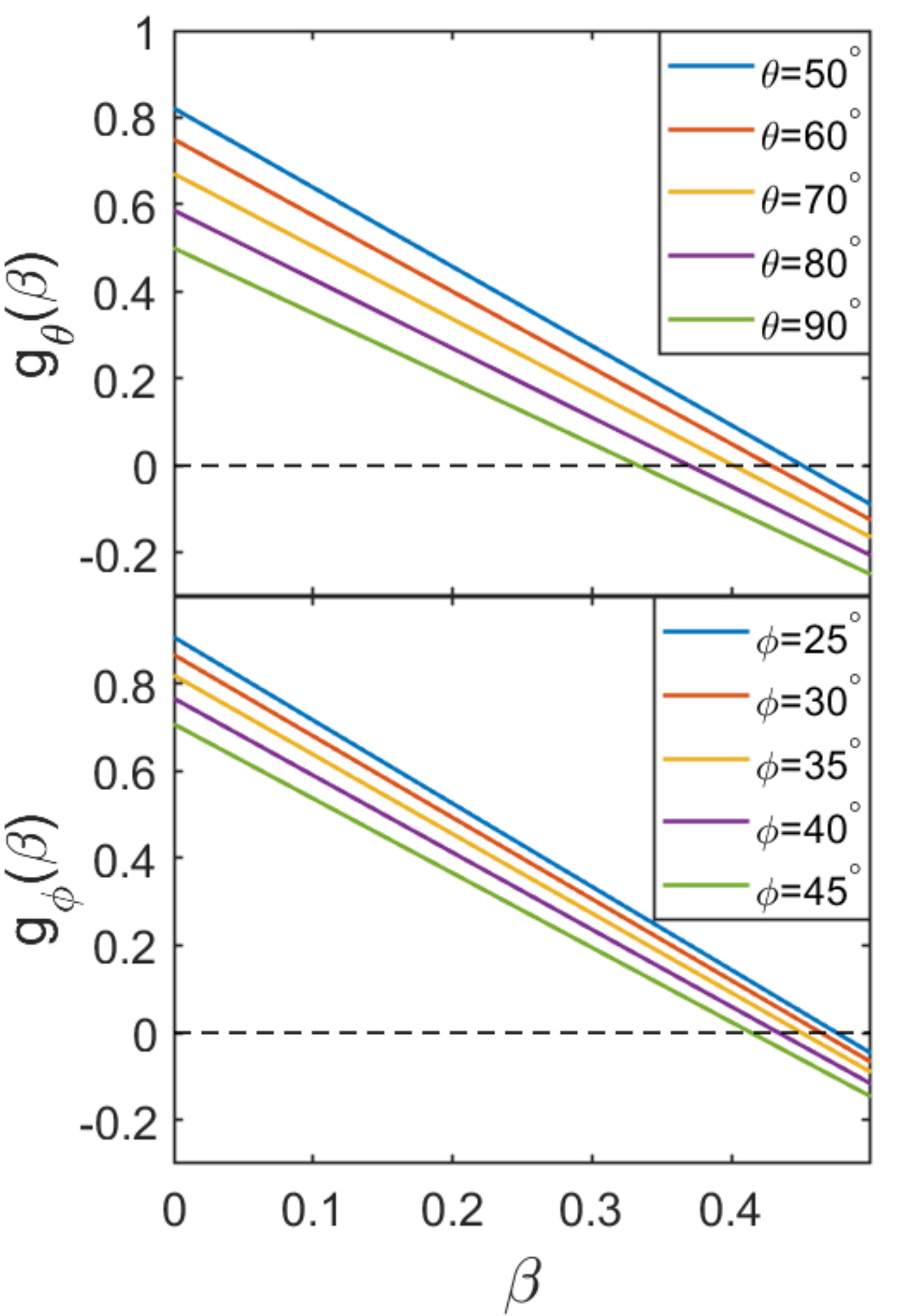}  
                 \vspace*{-25pt}
                 \caption{}
                 \vspace*{25pt}\label{fig:g_beta}
         \end{subfigure}%
    \begin{subfigure}[c]{0.37\textwidth}
                 \centering
                 \includegraphics[width=0.98\textwidth]{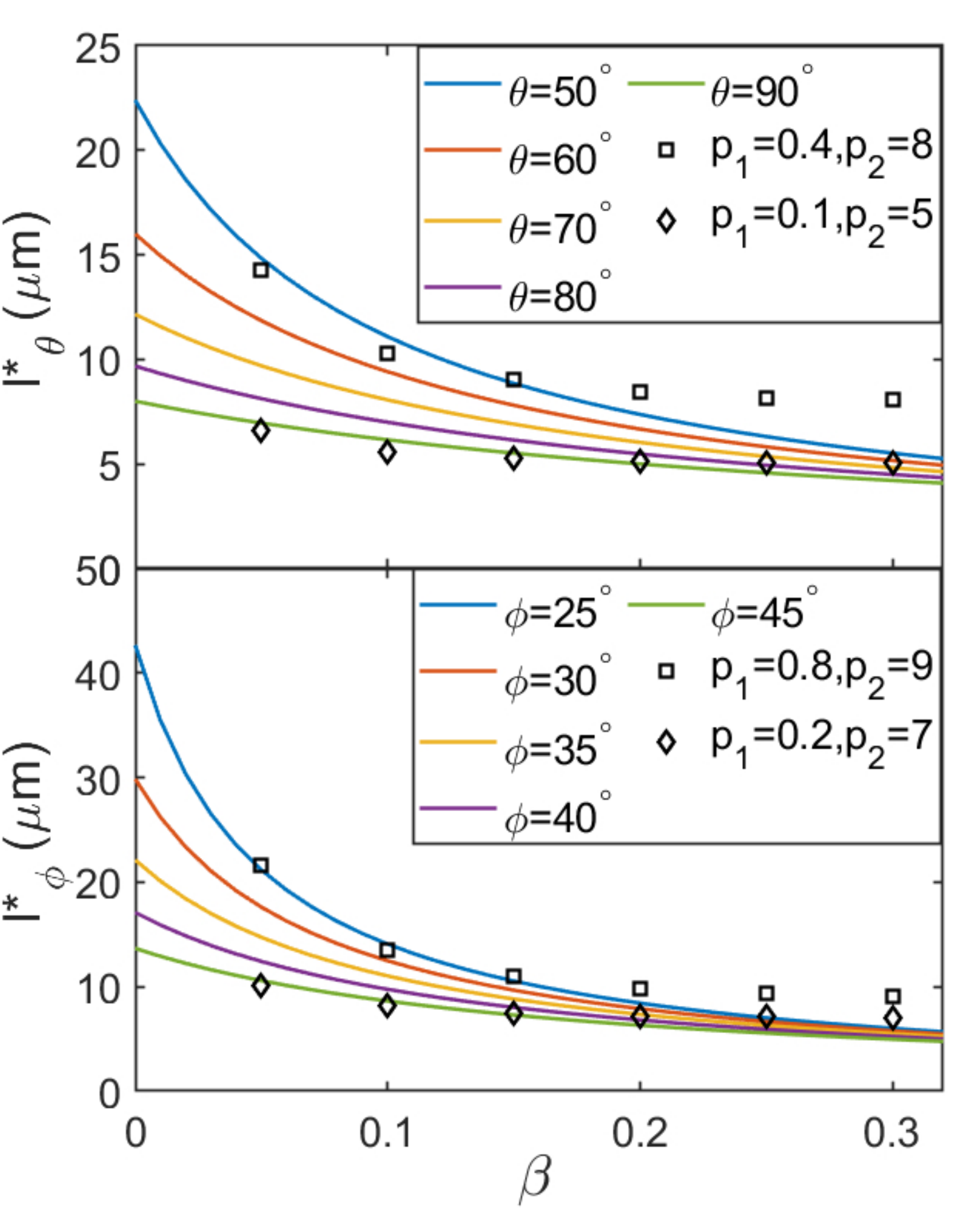}
                 \vspace*{-25pt}
                 \caption{}
                 \vspace*{25pt}
                 \label{fig:Lstar_beta}
         \end{subfigure}%
    \vspace*{-30pt}
    \caption{Scaling of transmission with size of the network, characterised by number of layers $N_L$ rather than physical size due to discreteness of networks. Averaged over 500 random realisations. (a) Linear scaling of $\langle \textrm{ln}(T_{total})\rangle$ gives a discrete version of localisation length, $\xi_N$. (b) Scaling of $\langle 1/T_{total}\rangle$. For smaller $\beta$ ($<$0.2), a linear trend is observed, following optical Ohm's law in the diffusive regime. For increased $\beta$ ($\geq$0.2), deviation from the linear trend indicate onset of localisation. Simulation results for  $\lambda=1500nm$ (crosses) and  $\lambda=750nm$ (circles) overlap due to the edge-length-invariant nature of scattering in the network system.  (c) Two simple models for a degree-four node described by Eq.~\ref{eq:abmatrix}, with the corresponding (d) scattering anisotropy $g$ and (e) transport mean free path $l^*$ shown for various scattering angles $\theta$ and $\phi$, calculated  using the similarity relation. Black markers in (e) are $l^*$ values computed from numerical scaling results as in (b), scaled by different parameters $p_1$ and $p_2$, defined in Eq.~\ref{eq:Lstar}. They deviate from the similarity relation results for $\beta\geq0.2$, i.e. when optical Ohm's law no longer holds.}
    \label{fig:scale_free}
\end{figure}

For localisation length, we define a size-invariant version $\xi_N$ such that $\xi=\xi_N d_L$, where $d_L$ is the average distance between layers in the grid network. This value is obtained via a linear fit to the scaling of $\langle \textrm{ln}(T_{total})\rangle$ with $N_L$, where the slope is taken to be $-1/\xi_N$ \cite{1DNecklaceStates,Doppler_2014}. The linear fit is performed for $N_L\geq 25$, where localisation has begun to set in. The scaling is shown for various backscattering strengths in Fig.~\ref{fig:lnTScaling}, with 
$\xi_N=34.43,29.32,24.28,20.10$ respectively for $\beta=0.1, 0.15, 0.2, 0.25$. As expected, increased level of backscattering leads to stronger localisation.

Fig.~\ref{fig:TinvScaling} shows the scaling of $\langle 1/T_{total}\rangle$ with $N_L$. A linear trend following optical Ohm's law is observed for weaker backscattering strengths, whereas larger $\beta$ deviate from this diffusive regime. Two sets of simulation are shown, one for $\lambda=1500~nm$ (crosses) and one for $\lambda=750~nm$ (circles), and both overlap. This is an illustration of the edge-length-invariance and ergodic property discussed above: since no wavelength-dependence is introduced at the scattering nodes, the statistical transmission properties over a random ensemble do not change with wavelength. 

In analogy with scattering in continuous random media, one would compute $l^*$ using the slope of the linear fit following optical Ohm's law. However, such a law is derived by solving the diffusion equation in finite size continuous media \cite{photonTMFP}, and it is not obvious how the method should be adapted for the discretised case on a planar network. Instead, we construct an embedded version of $l^*$ by relating it to the scattering mean free path $l_s$, which can be taken to be the average network edge length, and compare the results with the scaling data. $l^*$ and $l_s$ are related via the similarity relation $l^* = l_s/(1-g)$, where $g=\langle\textrm{cos}\varphi\rangle=\sum P(\varphi)\textrm{cos}\varphi$ is the anisotropy factor, averaging over all scattering angles weighted by their scattering probability $P(\varphi)$\cite{carminati_schotland_2021Book,Svensson2013SimilarityRelation}.

Fig.~\ref{fig:node_g_schemes} shows two simple schematics used to compute $g$ for the nodes described in Eq.~\ref{eq:abmatrix} and shown in Fig.~\ref{fig:Uscheme}. The first one treats a node simply as a cross, characterised by the angle $\theta$. In this case $g_\theta(\beta)=-\beta+\alpha(1-\beta)+\textrm{cos}\theta(1-\alpha)(1-\beta)$. Similarly, one may take into account the symmetry of a beamsplitter node and construct a node symmetrically with respect to the input port, characterised by the angle $\phi$, which gives $g_\phi(\beta)=-\beta+\textrm{cos}\phi(1-\beta)$. Setting $\alpha=0.5$ as before, Fig.~\ref{fig:g_beta} plots $g_\theta(\beta)$ and $g_\phi(\beta)$ for several different angles.  In both cases the node approaches isotropic scattering ($g=0$) for larger angles, or when $\beta$ is increased. Comparing with the scaling of $\langle 1/T_{total}\rangle$ in Fig.~\ref{fig:lnTScaling}, we see that the onset of localisation occurs at $\beta\approx 0.2$, which corresponds to roughly $g\approx0.5$ in the two node models.

The anisotropy is then plugged into the similarity relation to give $l^*_\theta$ and $l^*_\phi$, shown in Fig.~\ref{fig:Lstar_beta} as a function of $\beta$ for various angles. These values are compared with the simulation results for the scaling of $\langle 1/T_{total}\rangle$: even though the exact form of Ohm's law is unknown, we assume it takes the general form (in analogy to finite size continuum media \cite{3D_TMFP})
\begin{equation}
\label{eq:TvsL}
    T = \frac{c_1l^*+c_2}{N_L+c_3}
\end{equation}
where $c_i$ are constant parameters, and $N_L$ is used as the measure for network size instead of $L$. Then for each $\beta$, the inverse slope of the scaling shown in Fig.~\ref{fig:TinvScaling} (obtained via a linear fit) is linearly related to $l^*$:
\begin{equation}
\label{eq:Lstar}
    l^* = p_1\Bigl[\frac{\partial(1/T_{total})}{\partial N_L} \Bigr]^{-1}_\beta + p_2.
\end{equation}
By adjusting the fitting parameters $p_1$ and $p_2$, the resulting $l^*(\beta)$ follows $l^*_\theta$ and $l^*_\phi$ for different scattering angles, for small $\beta$ (Fig.~\ref{fig:TinvScaling}). The agreement between the two methods for computing $l^*$ fails for $\beta\geq 0.2$, where the transmission scaling deviates from optical Ohm's law, and is badly fitted by a linear trendline. In this sense, calculating $l^*$ using simple node models and the similarity relation captures the the effect of tuning $\beta$ on transmission scaling in the networks in the diffusive regime. Furthermore, if a network design has well-defined scattering angles at the nodes, this method provides a way to measure the phenomenological coefficients of optical Ohm's law, in a discretised planar setting. 

Due to their discretised nature and the size-invariant scattering behaviours, extra care needs to be taken if one wishes to treat the networks as direct analogues of other 2D multiple scattering systems. However, this feature also allows for flexibility in network design, since larger photonic devices can be used as scattering elements without altering the scattering behaviour of the overall network.

\section{Conclusion}
By modelling light propagating in integrated waveguides as a scalar wave, and treating the propagation and scattering separately, we have constructed a simple numerical model for multiple scattering of light in large scale photonic networks. In particular, the application to the LNOI platform has been experimentally verified to reach consistent agreement with multiple physical devices, providing a computationally inexpensive tool for simulating photonic networks and circuits. Furthermore, the model has been applied to a class of randomised grid-like networks, where we study multiple scattering phenomena in systems with well-defined type of randomness and scattering behaviour at each site. Such networks produce spectral transmission behaviour similar to that in other random media, while being tunable by the backscattering strengths at individual nodes. The networks are also unique in the discrete and statistically edge-length-invariant nature, which allows for more flexibility in terms of network designs and on-chip fabrication. By combining the computational simplicity of a graph-based network model and the control of top-down LNOI fabrication, these photonic networks provide a new on-chip platform for studying 2D multiple scattering with user-specified disorder and control. 
\section*{Acknowledgements}
We acknowledge support for characterisation of our samples from the Scientific Center of  
Optical and Electron Microscopy ScopeM and from the cleanroom facilities BRNC and  
FIRST of ETH Zürich and IBM Rüschlikon.  
X.S.W. acknowledges support from the ETH PhD fellowship. R.G. acknowledges support  
from the European Space Agency (Project Number 4000137426), the Swiss National Science  
Foundation under the Bridge Discovery Program LINIOS (Project Number 194693), and the  
European Research Council (Project Number 714837).

\section*{Disclosures}
The authors declare no conflicts of interest.

\clearpage

\end{document}